\RequirePackage{fix-cm}
\documentclass[smallextended]{svjour3}       % onecolumn (second format)
\smartqed  % flush right qed marks, e.g. at end of proof
\usepackage{graphicx}
\usepackage{color}
\usepackage[square,sort,comma,numbers]{natbib}
\usepackage{amsmath, amssymb, amsfonts}
\begin{document}
\sloppy

\title{Deep brain stimulation for movement disorder treatment: Exploring frequency-dependent efficacy in a computational network model}
%\subtitle{}

%\titlerunning{Deep brain stimulation for movement disorder treatment}        % 
\titlerunning{Exploring frequency-dependent efficacy in DBS}        % if too long for running head

\author{Konstantinos Spiliotis 
\and Jens Starke 
\and Denise Franz 
\and Angelika Richter
\and Rüdiger Köhling}

\institute{K. Spiliotis          \at
             Institute of Mathematics, University of Rostock, 
D-18057 Rostock, Germany \\
              \email{konstantinos.spiliotis@uni-rostock.de}           %  \\
           \and
           J. Starke \at
               Institute of Mathematics, University of Rostock, 
D-18057 Rostock, Germany \\
%              Tel.: +123-45-678910\\
%              Fax: +123-45-678910\\
              \email{jens.starke@uni-rostock.de}           %  \\
           \and
           D. Franz \at
           Oscar-Langendorff-Institute of Physiology,
Rostock University Medical Center, Rostock,
Germany\\
              \email{denise.franz@uni-rostock.de}           %  \\
           \and
           A. Richter \at
            Institute of Pharmacology, Pharmacy and Toxicology, Faculty of Veterinary Medicine, University of Leipzig,
Leipzig, Germany \\
            %  Tel.: +123-45-678910\\
             % Fax: +123-45-678910\\
              \email{angelika.richter@vetmed.uni-leipzig.de}           %  \\
            \and R. Köhling \at
            Oscar-Langendorff-Institute of Physiology,
Rostock University Medical Center, Rostock,
Germany\\
             % Fax: +123-45-678910\\
              \email{ruediger.koehling@uni-rostock.de}           %  \\
           \and
}

\date{Received: date / Accepted: date}

\maketitle

\begin{abstract}
A large scale computational model of the basal ganglia (BG) network is proposed to describes  movement disorder including deep brain stimulation (DBS). The model of this complex network considers four areas of the basal ganglia network: the subthalamic nucleus (STN) as target area of DBS, globus pallidus, both pars externa and pars interna (GPe-GPi), and the thalamus (THA). Parkinsonian conditions are simulated by assuming reduced dopaminergic input and corresponding pronounced inhibitory or disinhibited projections to GPe and GPi. Macroscopic quantities can be derived which correlate closely to thalamic responses and hence motor programme fidelity. It can be demonstrated that depending on different levels of striatal projections to the GPe and GPi, the dynamics of these macroscopic quantities switch from normal conditions to parkinsonian. Simulating DBS on the STN affects the dynamics of the entire network, increasing the thalamic activity to levels close to normal, while differing from both normal and parkinsonian dynamics. Using the mentioned macroscopic quantities, the model proposes optimal DBS frequency ranges above 130 Hz.
\end{abstract}

\section{Introduction}
\label{intro}
\subsection{Basal ganglia connectivity}
Parkinson's disease (PD) and dystonia, including different types, belong to the most common movement disorders, and hence pose a considerable health burden  \cite{Ches96,Def10,deLau06}.  
%(de Lau and Breteler, 2006; Defazio, 2010).
%Movement disorders such as Parkinson's disease and dystonias affect nearly 20 \% of the population worldwide \cite{INS16}, and hence pose a considerable health burden. 
It is generally accepted that they arise from a dysfunction of the basal ganglia  (BG), shown as  simplified circuitry in Fig.\ \ref{fig:1direct_inderect} and result in hypokinetic or hyperkinetic symptoms, depending on which part of the circuitry is affected. In this circuitry, in general, cortical, mainly glutamatergic projections are thought to activate GABAergic medium spiny neurons (MSN) and interneurones of corpus striatum (striatum in Fig.\ \ref{fig:1direct_inderect}). 
From the medium spiny, inhibitory neurones, the so-called direct pathway  projects to globus pallidus pars interna (GPi). 
The inhibition of GPi leads to activation of thalamus (via disinhibition); one can thus speculate that the thalamus more or less faithfully responds to initial cortical signals (to initiate movement). In the so-called indirect pathway, the activation of striatum inhibits the globus pallidus pars externa (GPe) which projects to the subthalamic nucleus (STN), enhancing its activity. An increased STN activation, in turn, will lead to an increment of GPi activity, resulting in an inhibition of the thalamus (and hence reduction of locomotive activity \cite{Cal14}), see also Fig.\ \ref{fig:1direct_inderect}. 
Parkinson's disease as a paradigmatic form of hypokinetic syndrome, due to degeneration of the substantia nigra (SN in Fig.\ \ref{fig:1direct_inderect}). It is characterised by rigidity, tremor and hypokinesia, i.e.\ the inability to start movements fluently \cite{DeLong90,Mai17}. Looking at the circuitry, the motor symptoms can reasonably well be explained by a loss of both activating and inhibitory projections of the SN to the striatum, leading to disinhibition in the so-called indirect (striatum - GPe - STN - GPi - THA), and over-activation in the so-called direct pathways (striatum - GPi - THA) \cite{Cal14}. Hyperkinetic syndromes, characterised by involuntary movements or muscle contractions, in turn, are generally thought to originate from functional or structural damage or degeneration of striatum (dystonia or choreatic syndromes) and of STN; ballistic syndromes. Again, the simplified circuitry depicted in Fig.\ \ref{fig:1direct_inderect} can serve to explain the functional outcome of e.g. striatal over-activation speculated to result in a shift of balance toward the so-called direct pathway in dystonias \citep{Wich11}. 
%Beyond what the circuitry model intuitively suggests, movement disorders can often be accompanied by non-motor symptoms such as depression or dementia, testifying to the fact that basal ganglia function also affects cognition and emotion \cite{AANS}.  \\
%The conditions include: spasticity; Parkinson’s disease; essential tremor; dystonia; Tourette syndrome; camptocormia; hemifacial spasm; and Meige syndrome.
% and  including chronic pain. While the underlying cause often cannot be cured, symptoms are managed through a multidisci plinary approach with the help of skilled healthcare providers.
%Treatment of movement disorders generally begins with oral medications, but if these are inadequate or produce unacceptable side effects, a number of neuromodulation approaches are available. For example, a neurostimulator device (similar to a heart pacemaker) can be implanted to deliver electrical stimulation to the brain, and this can improve symptoms of tremor, dystonia, or the stiffness and slowness seen in Parkinson’s disease. Alternatively, delivery of the anti-spasm agent baclofen into the spinal canal can improve spasticity.
 
\subsection{Role of activity patterns}
While the simplified circuitry suggests relatively straight-forward explanations for the emergence of hyper- or hypokinetic syndromes, in fact it does not take into consideration the patterning of activity - although information in the nervous system is actually conveyed by spatio-temporal activity patterns. Pattern propagation, however, will typically depend on non-linear couplings among the interacting compartments of the system \cite{Bev02}. Hence, one cannot assume that inhibitory or excitatory activity will straight-forwardly propagate across the network.  Importantly, both in PD and in dystonia, changes in oscillatory activity patterns in the basal ganglia and the cortex seem to be markers of the diseases \cite{Eus07}. Thus, in PD, synchronised beta-band activity in both cortex and STN seems to be associated with hypokinesia \cite{Cro12,Kuhn08}. This prominent beta-band synchronisation is speculated to be caused by a shift of network interactions among the different basal ganglia nuclei \cite{Schwab2013}. Indeed, a shift from complex spatio-temporal activity coupling between GPe and STN to strongly correlated and rhythmic patterns seems to underlie the pathophysiology of PD \cite{Bev02}. Synchronous abnormal patterns of activity of this local circuit should also be reflected in similar changes regarding the activity of the basal ganglia output and thalamic activity.

\subsection{Deep brain stimulation}
Deep brain stimulation (DBS) has been deemed to be the most important innovation in movement disorder therapy, and has revolutionised treatment for PD, dystonia and essential tremor patients, first having been approved for the latter condition by the FDA (Food and Drug Administration, USA) in 1997 \citep{Kra19}. Moreover, DBS is currently being introduced also for the therapy of mental disorders such as depression and obsessive-compulsive disorder \cite{Holtz11}. DBS clearly improves motors symptoms in PD, and often in dystonias \cite{Deus06}. For PD, we know that the efficiency of the treatment depends strongly on the frequency of the stimulus; high frequency stimulation (HFS), $f>90$ Hz \cite{McCon12} improves motor symptoms, while low frequencies are ineffective or worsen the clinical situation \cite{Koe19}.  
Until now, the main mechanism of DBS remains elusive \cite{Kra19,Ash17,Udu15}. The main hypotheses put forward so far are a) an inactivation of the target nuclei b) changes in transmitter release c) neuroprotective and electrostatic effects (i.e.\ structural plasticity), and network effects resulting in firing pattern alterations. While it seems clear that DBS does not cause a total silencing of neuronal activity of the target nuclei, but rather complex responses including sequences of prolonged activation / activity reduction \cite{Luo16} possibly because of a dissociation between somatal and axonal activation \cite{HOL00}. One of the most attractive explanations at the moment is the disruption of hypersynchronised oscillations \cite{Kuhn08}. Indeed, recordings from animal models suggest that DBS in the STN results in more periodic and regular firing at higher frequencies in the thalamus \cite{Xu08}. 

\subsection{Predicting network dynamics under DBS using computational approaches}
From a computational perspective, one of the main obstacles to explain or predict DBS effects on network activity is the lack of a coherent framework which could bridge the different scales \cite{Deco08,Siet16} of network models - ranging from microscopic (cellular activity) to macroscopic (symptom)\cite{Pav12,Pav15}. An intermediate level, the mesoscopic, is related with dynamics of specific networks of neurons in different nuclei of the basal ganglia. These mesoscopic networks  constitute the bridge between micro and macro scales. 
Examples of these network dynamics and variations of activity pattern are confirmed in a number of animal model studies in Parkinson models. Thus, DBS in the STN leaves firing rates unaltered in GPe, and increases firing rate in the thalamus (i.e.\ the nucleus the GPE projects on). While single unit recordings thus indicate changes in firing \emph{rates} in projection areas of the STN during DBS, one overarching motive of all observations is that DBS effectively changes firing patterns \cite{McCon12,Xu08,So12,so17,Dorv10} as essential element for therapeutic success. 

More specifically, in  \cite{McCon12}, DBS leads to reduced low-frequency neuronal oscillations, increased neuronal oscillations at the stimulation frequency, and increased phase locking with the stimulus pulses. Moreover, coherence within and across the GPe and SNr during HFS was reduced in the band of pathological low-frequency oscillations and increased in the stimulation frequency band. These findings provide evidence that effective high-frequency DBS suppresses low-frequency network oscillations and entrains
neurons in the basal ganglia. Therefore, these  results support the hypothesis that the effectiveness of HFS stems from its ability to override pathological firing patterns in the basal ganglia by inducing
a new regularized pattern of synchronous neuronal activity.

In this spirit we propose a computational large-scale biophysical model related to the Parkinson disease (PD) and DBS treatment. Based on the work presented in \cite{Ter02,Rub04} and using  complex network theory\cite{Bass06,Wat98,Bull09}, four areas of the BG; the globus pallidus (partes externa / interna)  (GPe-GPi), the subthalamic nucleus (STN) and the thalamus, are modelled. We show, in accordance with dopaminergic dysfunction occurring during Parkinson's disease,  how different levels of striatal inhibition to BG areas switch  the system dynamics from ``normal'' to a ``parkinsonian'', i.e. switching from faithful transfer of information through the network, to disturbed. The model can also reproduce the action of DBS in the STN, and discloses how high-frequency stimulation (HFS) influences the whole network in a computational view. Specifically, during DBS conditions, the model reveals a de-synchronisation or declustering of GPe and GPi activity which is projected to the thalamus. Defining and quantifying the response efficacy of thalamic activation during DBS, we deduce ranges of stimulation frequencies optimal for therapeutic success. Our model uses a significantly large number of neurons and connections (approx. two orders of magnitude larger compared to \cite{Ter02,Rub04,So12}) which makes the model more realistic and allows us to describe the behaviour of the neural network macroscopically. Although the detailed description is on the microscale (level of neurons), a macroscopic analysis is of interest, specifically the mean activity of interconnected neurons. The results of the macroscopic analysis suggests that a strong nonlinear response is obtained in the basal ganglia network, similar to resonating mechanical systems. The following analysis proposes optimal frequencies above 130Hz, suggesting the investigation of DBS treatment over 130Hz in animal experiments.

\begin{figure}
\begin{center}
\begin{picture}(270,250)
%\put(0,0)
\includegraphics[width=0.9\textwidth]{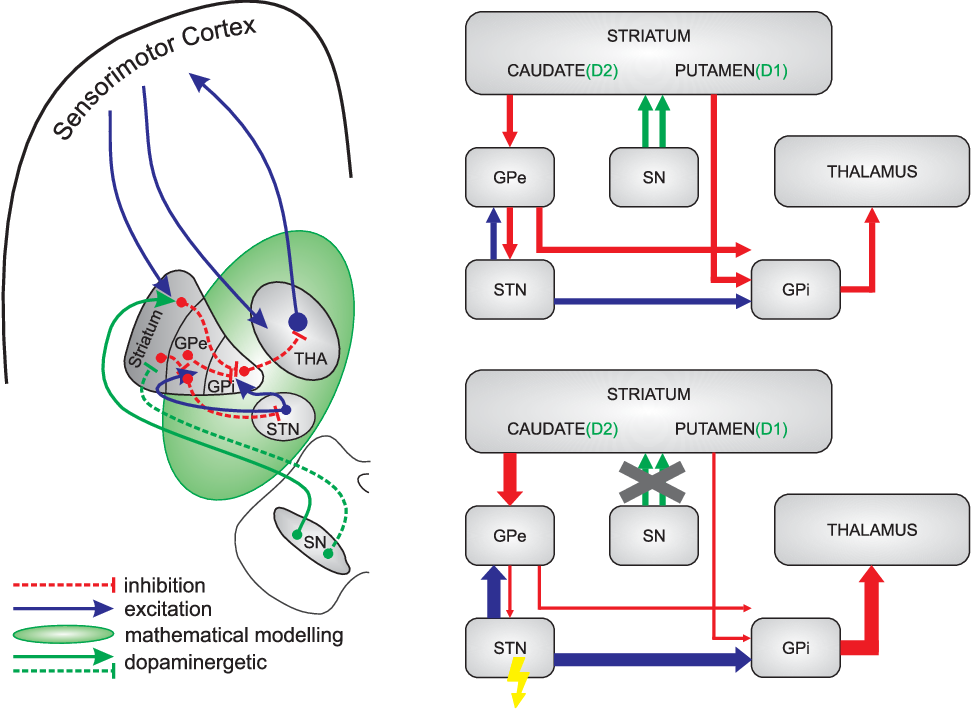}% This is a *.eps file
%}% This is a *.eps file
\put(-320,220){(a)}
\put(-180,220){(b)}
\put(-180,100){(c)}
\end{picture}
\end{center}
\caption{Representation of the basal ganglia network and the relevant connections in their approximate anatomical positions (a), as reduced functional scheme in healthy conditions (b) and under parkinsonian conditions (c), where the crossed out green arrows denote the functional loss of dopaminergic projection from substantia nigra. The sensorimotor cortex projects to the thalamus and to the striatum as input region of the basal ganglia. The basal ganglia network consists of the striatum, the globus pallidus, external and internal parts (GPe and GPi, respectively), the subthalamic nucleus (STN), substantia nigra (considering pars compacta here only, SN) and the thalamus. The green area in (a) marks the part of the network which has been modelled mathematically. Blue lines with arrows depict excitatory connections, while red lines inhibitory ones. Green arrows depict dopaminergic projections from the SN, which will activate the striatum in the direct pathway via D1 receptors, and inhibit the striatum in the indirect pathway via D2 receptors. In (b) and (c), bold lines depict increases in corresponding synaptic projections, and thin lines reductions, respectively. The thunderbolt arrow indicates the location of deep brain stimulation (DBS) in the STN. We hypothesize that under DBS conditions, the increased excitation (bold blue projection from STN to GPi) will be reduced, and the disinhibition (thin red projection from GPe to Gpi) will be normalized.}\label{fig:1direct_inderect}
\end{figure}
%For Original Research Articles \citep{conference}, Clinical Trial Articles \citep{article}, and Technology Reports \citep{patent}, the introduction should be succinct, with no subheadings \citep{book}. For Case Reports the Introduction should include symptoms at presentation \citep{chapter}, physical exams and lab results \citep{dataset}.

\section{Mathematical modelling of Basal Ganglia Neurons}
The model of the basal ganglia network consists of four interconnected areas the subthalamic nucleus (STN), the globus pallidus internal (GPi) and external (GPe) and thalamus (THA), see Fig.\ \ref{fig:1direct_inderect}. In this section we formulate the mathematical description for the neurons in each area of the basal ganglia (BG) and the thalamus. \\

%\subsection{Modelling and simulations of neurons in the subthalamic nucleus}
In the model, the STN plays a key role as DBS is applied there to treat Parkinsons's disease. The dynamics of each STN, GPe and GPi neuron  are governed by a Hodgkin Huxley formalism and the current balance equation for the membrane potential reads\cite{Ter02,Bev99,Popo19}:
\begin{gather}
C\frac{dV_i}{dt} =-I_{\text{LEAK}}-I_{\text{K}}-I_{\text{Na}}-I_{\text{Ca}}-I_{\text{T}}-I_{\text{AHP}}-I_{\text{syn}}+I_{\text{app}}+I_{\text{DBS}}\label{eq:HH_basal}\\%I_{\text{STST}}-I_{\text{GPST}}+I_{\text{DBS}}
 \frac{dx_i}{dt} =(x_{\infty}-x_i)/\tau_{x_i} \label{gate_STN_sol}\\
 \frac{d[\text{Ca}^{2+}]_i}{dt} =k_2\left(-I_{\text{Ca}}-I_{\text{T}}-k_{\text{Ca}}[\text{Ca}^{2+}]_i\right)
 \label{eq:Ca_STN},
\end{gather}
where $C$ is the membrane capacity, $V_i$ is the membrane potential of the $i$-th neuron, $x_i$ denote the gating variables $n,h,r$ and  $[\text{Ca}^{2+}]_i$ is the intracellular concentration of calcium. For all basal ganglia areas the currents are described below: The leak currents $I_{\text{LEAK}}=g_{\text{LEAK}}(V_i-E_{\text{LEAK}}),$ the potassium calcium and sodium  currents are given by $I_{\text{K}}=g_{\text{K}}n^4(V_i-E_{\text{K}})$, $I_{\text{Ca}}=g_{\text{Ca}}s^2_{\infty}(V_i-E_{\text{Ca}})$ and $I_{\text{Na}}=g_{\text{Na}}m^3_{\infty}h(V_i-E_{\text{Na}})$, while the low threshold T-type calcium current for STN is  $I_{\text{T}}=g_{\text{T}}a^3_{\infty}b^2_{\infty}(V_i-E_{\text{Ca}})$. In the case of GPe, GPi neurons, the low threshold calcium current has the form  $I_{\text{T}}=g_{\text{T}}a^3_{\infty}r(V_i-E_{\text{K}})$, reducing the bursting activity of the GPe compared to STN neurons. The current underlying the after-hyperpolarizing potential has the form $I_{\text{AHP}}=g_{\text{AHP}}([\text{Ca}^{2+}]/(k_1+[\text{Ca}^{2+}])(V_i-E_{\text{K}})$. 

The current $I_{\text{DBS}}$ in eq.\ \eqref{eq:HH_basal} models deep brain stimulation in STN neurons only and has the value $0$ without DBS treatment. The current $I_{\text{app}}$ is applied in STN, GPe and GPi, but with different physiological meaning. In the case of STN neurons, the $I_{\text{app}}$ simulating the afferent synaptic input from cortex\cite{Ter02}, while in the case of GPe and GPi the  $I_{\text{app}}$ represents the incoming signal from striatum with different levels of inhibition of GPe and GPi, respectively. 

In eq.\ \eqref{gate_STN_sol} the equilibrium state is $x_{\infty}=x_{\infty}(V_i)=1/(1+e^{-(V_i-\theta_x)/\sigma_x})$ for $ x=n,m,h,a,r,s$, while for  the equilibrium state  of T current the following form is used: $b_{\infty}(V_i)=1/(1+e^{(r_i-\theta_b)/\sigma_b})-1/(1+e^{-\theta_b/\sigma_b})$ The voltage dependent time scale $\tau_{x}$ has the form $\tau_{x}(V_i)=[\tau_{x0}+\tau_{x1}/(1+e^{-(V_i-\theta_{\tau x})/\sigma_{\tau x}})]/ A_x$, for STN and $\tau_{x}(V_i)=\tau$ for GPe and GPi \cite{Ter02}.

\begin{figure}
\begin{center}
\begin{picture}(340,220)
%\put(0,0)
\includegraphics[width=1.1\textwidth]{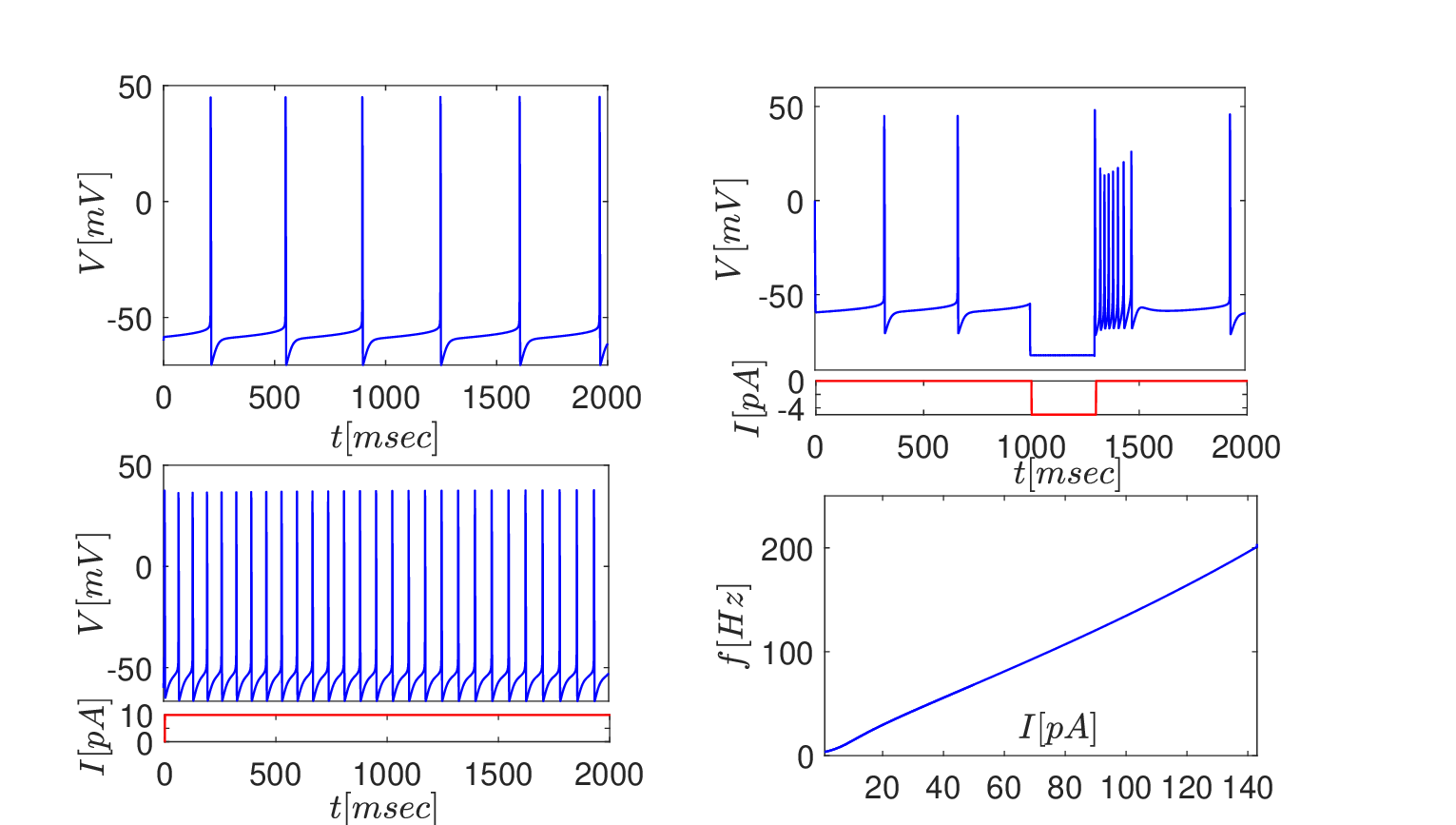}%}% This is a *.eps file
%\put(0,0)% This is a *.eps file
%}% This is a *.eps file
\put(-360,200){\textbf{(a)}}
\put(-200,200){\textbf{(b)}}
\put(-360,90){\textbf{(c)}}
\put(-200,90){\textbf{(d)}}
\end{picture}
%\put(0,6.5){(a)}
%\put(6.5,6.5){(b)}
%\put(0,3){(c)}
%\put(6.5,3){(d)}
%\end{picture}
\end{center}
\caption{Modelled activity of single STN neurons under different current injection input conditions described by eqs. \eqref{eq:HH_basal}, \eqref{gate_STN_sol}, \eqref{eq:Ca_STN}, and resulting current-frequency tuning curves. \textbf{(a)} STN neurons without current injection fire with a frequency around 3 Hz. \textbf{(b)} A negative current injection (current injection depicted in the lower part of the diagram, red curve) applied between t=1.0 and t=1.2 sec results in transient silencing of the neuron, and subsequent rebound firing due to $I_{h}$. \textbf{(c)} Injection of 10 pA positive current (current injection depicted in the lower part of the diagram, red curve) results in tonic firing activity at 15Hz. \textbf{(d)} Current-frequency tuning curve over the entire range of current injection modelled (0-150 pA). 
%Please note that the response is nonlinear and saturates at a firing frequency of $\approx 130$ Hz.
}
\label{fig:2STN}
\end{figure}
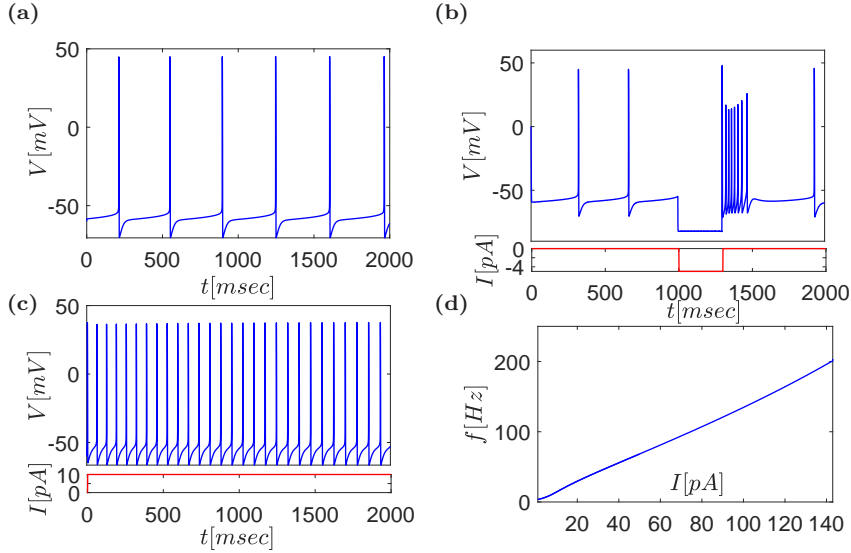

 Fig.\ \ref{fig:2STN} depicts the dynamics of one (uncoupled) STN neuron, firing at a frequency of 3Hz (all values of parameters are given in table \ref{tab:1}, see also \cite{Ter02}). When a negative current is applied for short time, the neuron is hyperpolarized accordingly. Due to the presence of hyperpolarization-activated currents (HCN currents), a rebound burst occurs after the current injection. 
 
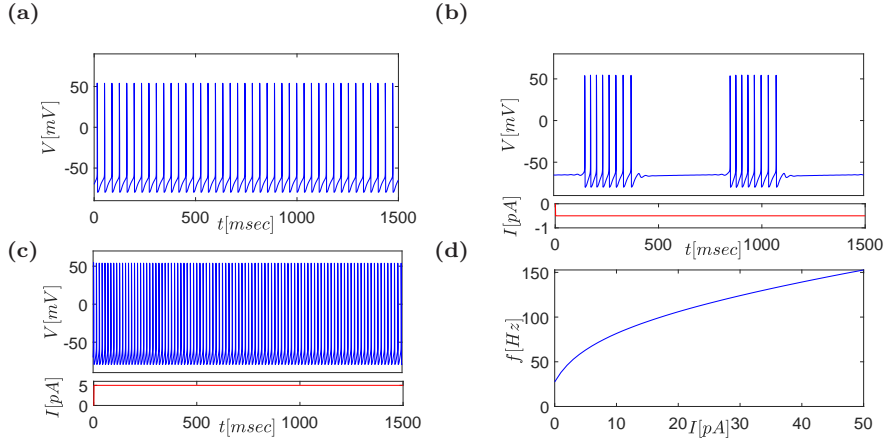
\begin{figure}
\begin{center}
\begin{picture}(380,180)
%\put(0,0)
\includegraphics[width=1.1\textwidth]{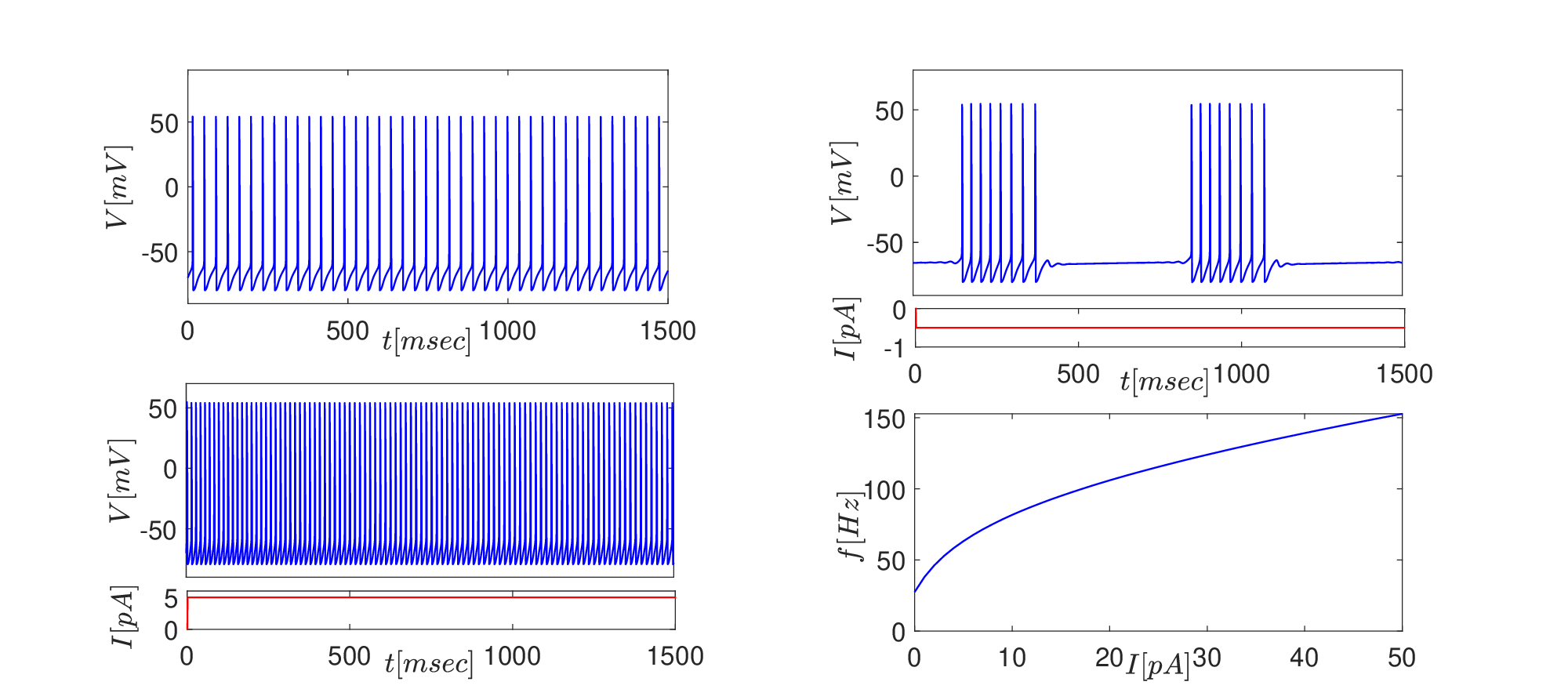}%}% This is a *.eps file
\put(-360,160){\textbf{(a)}}
\put(-200,160){\textbf{(b)}}
\put(-360,70){\textbf{(c)}}
\put(-200,70){\textbf{(d)}}
\end{picture}
%\put(0,6.5){(a)}
%\put(7,6.5){(b)}
%\put(0,3){(c)}
%\put(7,3){(d)}
%\end{picture}
%\includegraphics[width=14cm]{normal333.eps}% This is a *.eps file
\end{center}
\caption{Modelled activity of single GP neurons under different current injection input conditions described by eqs. \eqref{eq:HH_basal}, \eqref{gate_STN_sol}, \eqref{eq:Ca_STN}, and resulting current-frequency tuning curves. \textbf{(a)} GP neurons without current injection fire with a frequency around 30 Hz. \textbf{(b)} A constant negative current injection (current injection depicted in the lower part of the diagram, red curve) results in burst firing at 48 Hz. \textbf{(c)} Injection of 5 pA positive current (current injection depicted in the lower part of the diagram, red curve) results in tonic firing activity at 63 Hz. \textbf{(d)} Current-frequency tuning curve over the entire range of current injection modelled (0-50 pA).}
\label{fig:3GPE}
\end{figure}

According to experimental findings \cite{Ple99,kita91},  GP neurons are characterized in contrast to STN cells with similar ionic currents but different parameters and the resulting dynamical properties are similar to experimental data in \cite{Ple99,kita91}. The main attributes are a spontaneous firing activity at a frequency of $\approx$ 30Hz  \cite{Coop00,kita91}, (see also \cite{Ter02} and the references there in), and a rebound response to a  hyperpolarizing current \cite{Ter02,kita91,Coop00}. Simulations of one GPe/GPi neuron are shown in Fig.\ \ref{fig:3GPE}, where the GP neuron fires at a frequency of 30Hz at rest shown in subfigure \ref{fig:3GPE}(a). Fig.\ \ref{fig:3GPE}(b) depicts the dynamics for small negative $I_{app}$ resulting in intermittent bursting activity. A small depolarizing current, in turn, results in fast tonic discharges at nearly twice the resting discharge frequency \ref{fig:3GPE}(c). The entire tuning curve depicting responses to different levels of current injections is depicted in \ref{fig:3GPE}(d). 
 
 \subsection{Description of the basal ganglia synaptic connectivity}
 \label{sec:STN_conn}
The coupling between the neurons is described by the synaptic current $I_{\text{syn}}$.  
In the model, GPe and GPi neurons are connected through a Watts and Strogatz (WS) small-world topology \cite{Wat98,Bull09,Stam2007,Gaf16,Sant14,Net04,Ber17,Fang17}, where neurons nearby are densely connected, additionally a small number of large distance connections exist. Following the experimental findings \cite{Gout18} which suggest sparse connections in the STN area, we chose a modified small world topology with substantially  smaller intra-striatal connectivity. The detailed description of the network structure is given in sec.\ \ref{sec:Network}.

The small-world network is considered in the synaptic currents defined by the activation variable $s$, which are given by \cite{Lai02,Erm12,Com09}:
\begin{equation}
\frac{ds_i}{dt} = \alpha(1-s_i)H(V_i-\theta_{0})-\beta s_i ,
\label{eq:synaptic_Gen}
\end{equation}
where $H(V)$ is a smooth approximation of step function, i.e. $H(V)=1/(1+e^{-(V-\theta_x)/\sigma_x}$.

Typically, fast synapses are of order $\alpha, \beta  = O\left( 1 \right)$ 
while for slow synapses holds $\alpha=O\left(1 \right)$ and 
$\beta  = O\left(\epsilon \right)$, meaning that fast synapses activate and deactivate following fast time scales, while the slow synapses activate fast, and deactivate slowly \cite{Lai02,Erm12,Term05}. 

The excitatory and inhibitory synaptic currents for the $i$-th neuron are given, respectively, by
\begin{equation}
I_{i,\text{Glu}}=g_{\text{Glu}}(V_i-E_{\text{Glu}})\sum_{j}{A_{ij}s_{j}}, 
\label{eq:GluI}
\end{equation}
with $E_{\text{Glu}}=-10mV$, and
\begin{equation}
I_{i,\text{GABA}}=g_{\text{GABA}}(V_i-E_{\text{GABA}})\sum_{j}{A_{ij}s_{j}},
\label{eq:GabbaI}
\end{equation}
with $E_{\text{GABA}}=-70mV$.
where $A_{ij}$ has the value 1 or 0, depending on whether the neuron is connected or not. The summation is taken over all presynaptic neurons.

In the case of STN neurons the $I_{\text{syn}}$ current, is given by the summation  $I_{\text{syn}}=I_{\text{STST}}+I_{\text{GPST}}$, indicates the internal excitation between the STN neurons and the incoming inhibition from the GPe neurons, respectively. The excitatory glutaminergic  connections within the STN are expressed by  $I_{STST}$ which follows the eq.\ \eqref{eq:GluI}, while the inhibitory current $I_{GPST}$ is given by eq.\ \eqref{eq:GabbaI} and express the inhibition from the GPe area.

The synaptic current $I_{\text{syn}}$ for the GPe region is defined by $I_{\text{syn}}=I_{\text{GPeGPe}}+I_{\text{STGPe}}$, where the first term $I_{\text{GPeGPe}}$ express the interlayer inhibitory interaction of GPe neurons (i.e. follows eq.\ \eqref{eq:GabbaI}), while $I_{\text{STGPe}}$  describes excitation from STN neurons. For the GPi region the current $I_{\text{syn}}$ is given by $I_{\text{syn}}=I_{\text{GPiGPi}}+I_{\text{GPeGPi}}+I_{\text{STGPi}}$, where the first two terms $I_{\text{GPiGPi}}$ and $I_{\text{GPeGPi}}$ are inhibitory  connections, connections from GPi to itself and from GPe to GPi, respectively, while $I_{\text{STGPi}}$  describes excitations from STN neurons. The values of the parameters are given in table \ref{tab:1}, see also tables 1, 2 of \cite{Ter02}.

%\subsection{Level 2}
%\subsubsection{Level 3}
%\paragraph{Level 4}
%\subparagraph{Level 5}

\subsection{Modelling and simulations of neurons in the thalamus}

\begin{figure}
\begin{center}
\includegraphics[width=1.\textwidth]{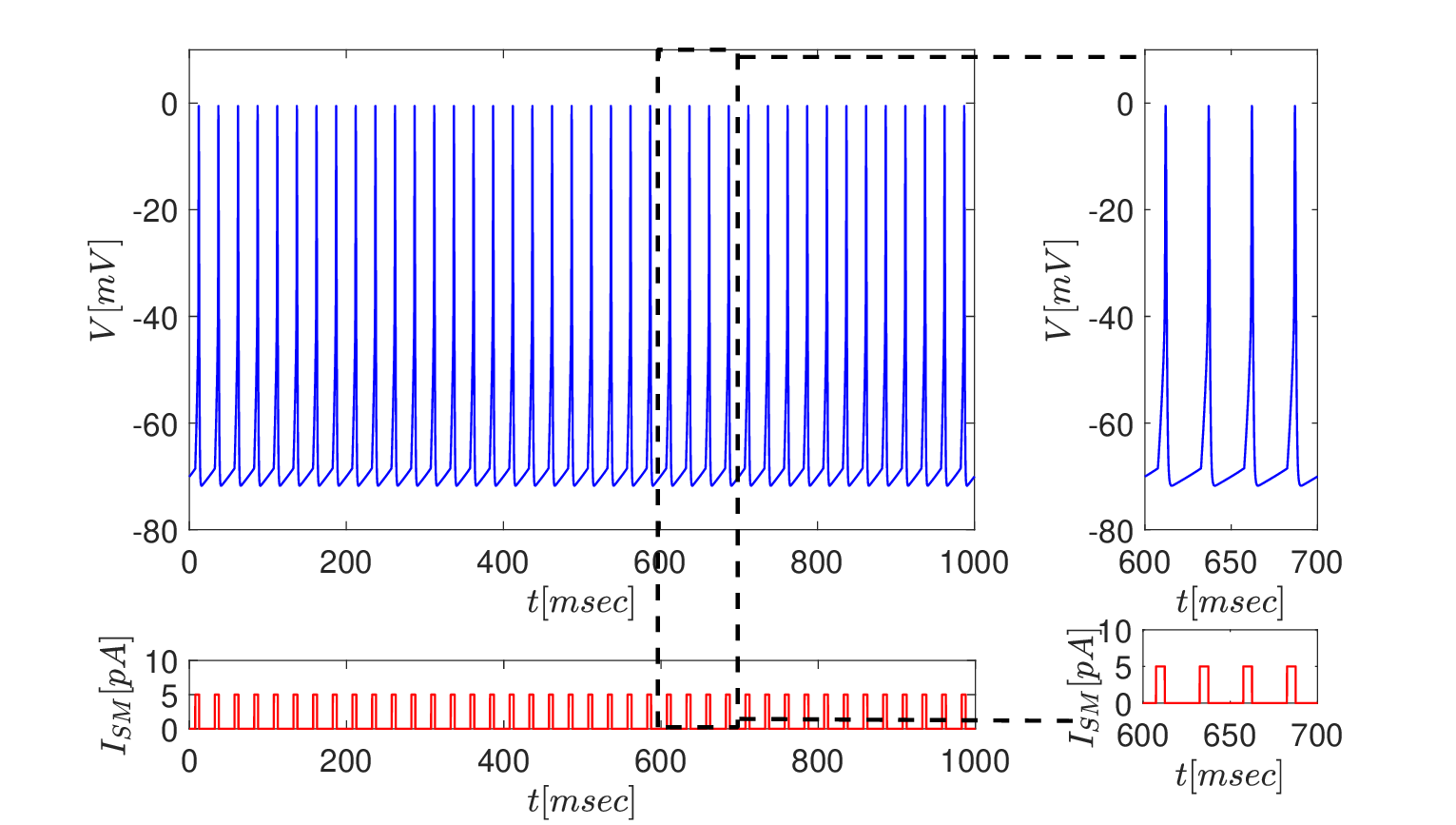}% This is a *.eps file
\end{center}
\caption{Firing of a single thalamic neuron receiving periodic sensorimotor cortex input current $I_{SM}$. Cortical input was simulated by periodic 5 pA, 5 ms current injections (red curve) and corresponding membrane potential changes were calculated (blue curve). Isolated thalamic neurons faithfully follow the external periodic stimulus delivered at 40 Hz in this example (see also inset).}
\label{fig:4Thalam}
\end{figure}

Modelling the basal ganglia network, the crucial behaviour of the model is the response of thalamic neurons to synaptic input from GPi neurons. The following section hence addresses this question. Here, a movement-programme associated sensory-motor cortex input to the thalamus is simulated by a repetitive periodic synaptic activation. This is modelled by 5 pA, 5 ms current injections, at 40 Hz (as depicted in Fig.\ \ref{fig:4Thalam}), of the form \cite{Rub04}: 
$ I_{\text{SM}}=A_{\text{SM}}H(\sin(2\pi t/T_{\text{SM}})\cdot (1-H(\sin(2\pi(t+\delta_{\text{SM}})/T_{\text{SM}}))$, combined with GP neuronal input (resulting, in turn, from basal ganglia network activity) as main parameters determining thalamic firing.  

The mathematical description of thalamic neurons is given in the following equation
\begin{equation}
C\frac{dV_i}{dt}=-I_{\text{LEAK}}-I_{\text{K}}-I_{\text{Na}}-I_{\text{T}}-I_{\text{GPTH}}+I_{\text{SM}} ,
\end{equation}
where $C$ is the membrane capacity and $V_i$ is the membrane potential of the $i$-th neuron. The leak current has the form  $I_{\text{LEAK}}=g_{\text{LEAK}}(V_i-E_{\text{LEAK}}),$ the other ionic currents, i.e. potassium and sodium, are given by $I_{\text{K}}=g_{\text{K}}(0.75(1-h))^4(V_i-E_{\text{K}})$ and  $I_{\text{Na}}=g_{\text{Na}}m^3_{\infty}h(V_i-E_{\text{Na}})$, while the low threshold T type calcium current is described by   $I_{\text{T}}=g_{\text{T}}p^2_{\infty}r(V_i-E_{\text{T}})$.
The gating variables $h,r$ follow the differential equation with first order kinetics as in eq.\ \eqref{gate_STN_sol}
\begin{equation}
    \frac{dx}{dt}=(x_{\infty}-x)/\tau_{x} .
\end{equation}
The equilibrium function has the form $x
 _{\infty}(V_i)=1/(1+e^{-(V_i-\theta_x)/\sigma_x})$ for $x=r,h,m,p$ and the voltage dependent time scale for the gating variable $r$ is $\tau_{r}=28+(1+e^{-(V_i+25)/10.5})$, while for $h$  is defined as $\tau_{h}=1/(a_h+b_h)$ with $a_h=0.128e^{-(V_i+46)/18}$ and $b_{h}=4/(1+e^{-(V_i+23)/5})$. The current $I_{\text{GPTH}}$ represents the inhibition of GPi area to thalamus and has the form of eq.\ \eqref{eq:GabbaI}. For detailed description of thalamic neurons and for the arithmetic values of parameters, see \cite{Rub04} and table \ref{tab:2}.
 
As Fig.\ \ref{fig:4Thalam} shows, within an isolated cortico-thalamic interaction, the thalamus follows cortical input absolutely faithfully. The next obvious question is how the basal ganglia network input will modify this strictly coupled cortico-thalamic interplay.

%Frontiers requires figures to be submitted individually, in the same order as they are referred to in the manuscript. Figures will then be automatically embedded at the bottom of the submitted manuscript. Kindly ensure that each table and figure is mentioned in the text and in numerical order. Figures must be of sufficient resolution for publication \href{http://home.frontiersin.org/about/author-guidelines#ResolutionRequirements}{see here for examples and minimum requirements}. Figures which are not according to the guidelines will cause substantial delay during the production process. Please see \href{http://home.frontiersin.org/about/author-guidelines#GeneralStyleGuidelinesforFigures}{here} for full figure guidelines. Cite figures with subfigures as figure \ref{fig:2}B.
%\subsubsection{Permission to Reuse and Copyright}
%Figures, tables, and images will be published under a Creative Commons CC-BY licence and permission must be obtained for use of copyrighted material from other sources (including re-published/adapted/modified/partial figures and images from the internet). It is the responsibility of the authors to acquire the licenses, to follow any citation instructions requested by third-party rights holders, and cover any supplementary charges.
%%Figures, tables, and images will be published under a Creative Commons CC-BY licence and permission must be obtained for use of copyrighted material from other sources (including re-published/adapted/modified/partial figures and images from the internet). It is the responsibility of the authors to acquire the licenses, to follow any citation instructions requested by third-party rights holders, and cover any supplementary charges.

\subsection{Network structure}
\label{sec:Network}
\begin{figure}
\begin{center}
\includegraphics[width=1.2\textwidth]{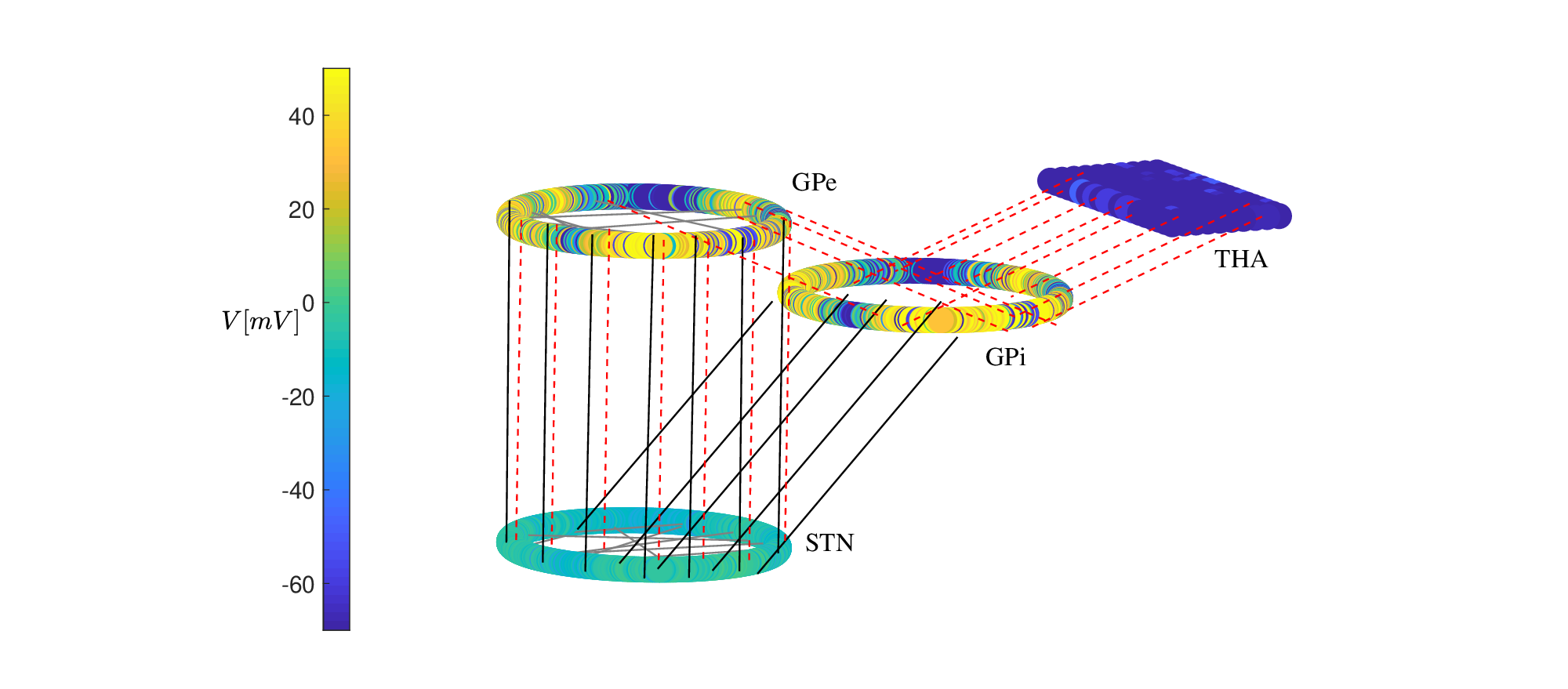}% This is a *.eps file
\end{center}
\caption{Schematic representation of the basal ganglia network and its activity pattern (voltages of membranes neurons in color). Black solid lines represent excitatory inputs from STN. Red dotted lines represent inhibitory connections from GPe to STN and GPi, and from GPi to TH, respectively. The GPe and GPi are considered to have a small world network structure, i.e. also contain sparse long-range connections (black horizontal lines), while the STN follows the same structure with reduced number of connections. Each GPe neuron, in turn, is linked with one STN neuron, and vice versa. Furthermore, each STN neuron is connected to one GPi neuron. For the model 500 STN neurons, 500 GPe neurons, 500 GPi neurons and 200 Thalamic neurons were used. A  total of 200 GPi neurons were uniformly random chosen to be connected to thalamic neurons. The ring structures represent relative structural neighbourhood relationships emphasised for GP and STN.}
\label{fig:5net}
\end{figure}

For modelling and analysis $N_{\text{STN}}=500$ STN neurons, $N_{\text{GPe}}=500$ GPe neurons, $N_{\text{GPi}}=500$ GPi neurons were used, while the thalamus was represented by $N_{\text{THA}}=200$ neurons. Following \cite{Wat98,Bass06,Bull09,Stam2007,Spil11}, the GPe/GPi layers were modelled as separate small world networks i.e. the connections of neurons follow a small-world topology. In such small world complex  networks \cite{Mark03}, not only does each neuron (node) in the network interact with its $k$ nearest neighbours, but there are also few randomly chosen  remote connections \cite{Wat98}. For the construction of nuclei networks (GPe and GPi) the value $k=20$ was used, while remote connections were created according to \cite{Wat98} with $p=0.005$. Fig.\ \ref{fig:5net} shows a characteristic snapshot of the network. For the STN, we chose a modified approach with reduced intra-striatal connectivity, as it has been already mentioned in sec.\ \ref{sec:STN_conn}, matching experimental findings which suggest sparse connections \cite{Gout18,Amm10}. Specifically, 80\% of STN neurons do not have any coupling and only the 20\% of the remaining STN neurons having collaterals within the other STN neurons \cite{Gout18}. For these neurons the distance of axonal endings with collaterals within the STN, showing that  roughly 30\% of all synapses lie within a 200 µm radius, and another 45\% within the 200-400 µm radius. The other 20\% are contacts which occur farther away, i.e.\ $>500$ µm. In this sense, the 20\% of neurons, which form STN connections, have both local and remote connections similar to small world property (i.e.\ 20\% of neurons only showing an average of 25 connections each \cite{Gout18}). Finally, the case of denser connectivity between STN neuron is also studied in section \ref{sec:dense} and the results are shown in Fig.\ \ref{fig:travel_welle2}.

The coupling between the layers is achieved in the following way: Each GPe neuron is connected to one STN neuron and vice versa. Furthermore, each STN neuron is communicating with one GPi neuron. In addition to excitatory inputs, the GPe also sends inhibitory signals to the  GPi according to small world topology, see Fig.\ \ref{fig:5net}. 
Each thalamic neuron, in turn, is receiving inhibitory inputs from three GPi neurons. Since the main interest of this study focuses on the impact of the basal-ganglia output activity on thalamic response to sensorimotor signals  (shaping the thalamocortical interplay), this model does not include \emph{intrathalamic} connections.

%We consider three states of connectivity strength of the layers in the basal ganglia. During the normal state we impose low inhibition  from GPi to thalamus. This has a minimal effect on the thalamic cells.
%. Finally, we simulate DBS of STN neurons. We
%assume that DBS provides a high frequency, excitatory
%input to STN neurons. We find that this input leads to
%increased activity of STN neurons which in turn excite GPi cells, inducing them to fire tonically at high frequency. Our main result is that this can restore the ability of the thalamus to relay its sensorimotor input faithfully

\section{Macroscopic description of basal ganglia dynamics}
\subsection{Synchronisation analysis of basal ganglia dynamics}  \label{sec:synch}

In the following, a nonlinear dynamical system of the form
\begin{equation}
    \dot x=f(x), \quad x\in \mathbb{R}^n,
    \label{eq:System_gen}
\end{equation}
is considered, where $f$ is a nonlinear vector field. 
%If a solution $x=x(t)$ converges to a constant value $x=x^* \in \mathbb{R}^n$ this is called a stationary state. 
For a periodic (or oscillatory) solution the property
\begin{equation}
    x(t+T)=x(t),
    \label{eq:limit_C}
\end{equation}
holds with period $T$, where $x(t)$ defines a periodic orbit in the phase space. 
If the periodic orbit shows normal hyperbolicity \cite{Izhi04}, the system of eqs.\ (\ref{eq:System_gen}) can be transformed into an angle or phase equation for the angle $\theta$, i.e. 
\begin{equation}
      \dot \theta = \omega,
      \label{eq:theta}
\end{equation}
where $\omega=2\pi/T$ is the natural frequency of the oscillator.

Coupled oscillators described as system of differential equations (\ref{eq:System_gen}) are of particular interest in neuroscience. Under certain conditions, identical or nearly identical interacting oscillators converge to a common frequency and synchronise \cite{Str01}. If the coupling strength is small, then each oscillator stays almost at its natural frequency. Increasing the coupling strength beyond a specific threshold, a group or cluster of oscillators shows synchronisation properties \cite{Str01}.

For two coupled oscillators, this phase-locking is expressed as:
\begin{equation}
   \exists m,n \in \mathbb{Z} : m\theta_1(t) -n\theta_2(t)=c, \quad \forall t .
\end{equation}
The case $m=n=1$ and $c=0$ or $c=\pi$  corresponds to phase and antiphase synchronisation, respectively. The appearance of synchronisation between two oscillators depends on two crucial parameters: The distance between the natural frequencies $d=\omega_1-\omega_2$ and and the coupling strength $\epsilon$. High values of frequency distance $d$ decrease the synchronisation, while higher coupling strengths tend to synchronise the oscillators. In the two-parameter space $(d, \epsilon)$, the areas where the phase locking, i.e., the synchronisation, appears as `Arnold’ tongues \cite{Low16}.

The weak coupling of $n$ oscillators \cite{Izhi04} is formulated by 
\begin{equation}
    \dot x_i=f_i(x_i)+\epsilon g_i(x_1,x_2,..,x_n)=f_i(x_i)+\epsilon \sum_{j=1}^{n}g_{ij}(x_i,x_j) ,
    \label{eq:System_gen_Coupl}
\end{equation}
where $\epsilon\ll 1$ represents the coupling strength. The summation is over the coupling between $j$ and $i$, namely the adjacent of the $i$ oscillators. Similarly, there is a transformation which allows to express the dynamics in phase variables, i.e.
\begin{align}
     \dot \theta_i & = \omega_i+Q(\theta_i)\epsilon h_i(\theta_1,\theta_2,...,\theta_n)\\
     & = \omega_i+\epsilon \sum_{j=1}^{n}h_{ij}(\theta_i,\theta_j)
    \label{eq:System_gen_Coupl_theta}
\end{align}
where $\theta_i \in [0,2\pi]$. The weakly coupled system in phase form, i.e.\ eq.\ (\ref{eq:System_gen_Coupl_theta}), shows a phase-locked solution if there is a constant integer matrix $K_{n-1,n}$, such that $K\cdot \theta=c$ with $c_k>0$ \cite{Izhi04}. Furthermore, the coupling oscillators are synchronised (in-phase) when
\begin{equation}
  \theta_1(t) = \theta_2(t)=... = \theta_n(t)=\theta(t) .
  \label{eq:Synch}
\end{equation}
%For a stability analysis of the synchronised solution, the phase perturbation $\phi_i$ i.e.
%\begin{equation}
 %   \phi_i=\theta_i-\omega_i t 
%\end{equation}
%is introduced. Eq.\ (\ref{eq:System_gen_Coupl_theta}) can then be written as 
%\begin{equation}
%     \dot \phi_i  = \epsilon \sum_{j=1}^{n}h_{ij}(\omega_it+\phi_i,\omega_jt+\phi_j) .
%    \label{eq:System_gen_Coupl_phi_Aver1}
%\end{equation}
%By averaging \cite{Sand07}, the expression of eq.\ (\ref{eq:System_gen_Coupl_theta}) results in 
%\begin{equation}
%     \dot \phi_i  = \epsilon \sum_{j=1}^{n}H_{ij}(\phi_i-\phi_j) ,
%    \label{eq:System_gen_Coupl_phi_Aver2}
%\end{equation}
%with the function $H$ after averaging is given by
%\begin{equation}
%    H_{ij}(y)=\lim_{T\Rightarrow \infty} 1/T\int_{0}^{T}h_{ij}(\omega_it,\omega_jt-y)dt .
%    \label{eq:System_gen_Coupl_phi_Aver3}
%\end{equation}
%The stability of the synchronized solution, depending on the properties of function $H$, in some cases can be determined by using \cite{Izhi97,Erme92}. As no fully synchronized solutions emerge in the present study, the above theory cannot be applied.\\
Following the approach of Kuramoto \cite{ARENAS08,Kur84,Kur84B,Ber17} which is applied when oscillators are near supercritical Andronov-Hopf bifurcation, in case of a fully connected network eq.\ (\ref{eq:System_gen_Coupl_theta}), we obtain 
\begin{equation}
     \dot \theta_i = \omega_i+k/N\sum_{j=1}^{N} \sin(\theta_i-\theta_j),
\end{equation}
while with arbitrary connectivity i.e. a complex network topology
\begin{equation}
     \dot \theta_i = \omega_i+k/N\sum_{j=1}^{N}A_{ij}\sin(\theta_i-\theta_j) ,
\end{equation}
where the coefficient $A_{ij}\in \{0,1\}$ is derived from the network adjacency matrix \cite{Ber17}. 
Immediately, the macroscopic emergent dynamics (in terms of phase) can be obtained by taking the mean value of all phase populations (in exponential form, $e^{i\theta}$). This defines the \emph{synchronisation index $r$}, i.e. \cite{Kur84,Str01}
\begin{equation}
r(t)=\left| \frac{1}{N}\sum_{k=1}^{N}e^{i\theta_k(t)}\right| 
    \label{eq:Sync_index}
\end{equation}
This index is used in the following sections as a measure describing the level of synchronised patterns within the GPi. 
\\
The synchronisation index acts as order parameter with range $r\in[0,1]$. In the case of perfect synchronisation, i.e.\ eq.\ (\ref{eq:Synch}), the index can be written
 \begin{equation}
    r=\left| \frac{1}{N}\sum_{k=1}^{N}e^{i\theta_k}\right|=\frac{1}{N}N \left|e^{i\theta}\right|=1 , 
\end{equation}
while the case of $r \rightarrow 0$ corresponds to incoherent phase movement. The phase $\theta_k(t)$ used in eq.\ \eqref{eq:Sync_index} of the $k$-th neuron can be approximated linearly according to the following equation
\begin{equation}
    \theta_k(t)=2\pi \frac{t-t_n}{t_{n+1}-t_n}+2\pi n ,
    \label{lineartheta}
\end{equation}
where, $t_n$ corresponds to $n$-th firing time of the $k$-th neuron. Another option to compute the phase without assuming a linear  dependence of the angle $\theta$ on time, see eq.\ \eqref{lineartheta} is the Hilbert transform, applied by Gabor in \cite{Gab46}.  For a given function $x=x(t)$, the Hilbert transform is defined as
\begin{equation}
    X(t)=\frac{1}{\pi}P\int_{-\infty}^{\infty}\frac{x(\tau)}{t-t}d\tau
    \label{eq:Hilbert}
\end{equation}
with $P$ denoting the Cauchy principal value. The complex signal $z=z(t)=x(t)+iX(t)=A(t)e^{i\theta(t)}$ is defined in order to extract the phase, where $A, \theta$ is the amplitude and the phase of the complex signal $z$, respectively. The instantaneous phase is defined as:
\begin{equation}
  \theta(t)=\arctan{(X(t)/x(t))} .
\end{equation}
The synchronisation index eq.\ \eqref{eq:Sync_index} will be computed in the next section to characterise the network activity for the  cases of normal, parkinsonian and DBS treatment. 

An alternative macroscopic quantity which can be used to estimate neuronal activity is the \emph{mean synaptic activity index $l$}. It is defined by
\begin{equation}
    l(t)=\frac{1}{N}\sum_{i=1}^{N}s_i(t) ,
    \label{eq:LFP}
\end{equation}
where $s_i$ is the synaptic variable as it is defined from eq.\ \eqref{eq:synaptic_Gen}. 
% n vs N????n
Coupling theoretical modelling tasks with experimental data, the synchronisation property in essence reflects local field potentials (LFP) in the sense that the flow of the extracellular current generating the LFP representing summed postsynaptic potentials from
local cell groups (similar to eq.\ \eqref{eq:LFP}, see \cite{Buz04,Popo19,Manos18, Popo18}). In the next section this mean synaptic activity will be used to measure the transfer of neuronal activity (current flow information) from the basal ganglia output, i.e., from GPi to thalamus.

\section{Simulating different functional states of the basal ganglia network}

Three cases of dynamic network behaviours were studied. In all cases the thalamus receives a periodic sensorimotor input (simulated by periodic 5 pA current injections \cite{Rub04}) which represents the signal for the initiation of movement. In addition, a continuous input current to STN is applied in order to obtain rhythmic activity (simulating afferent synaptic input from cortex \cite{Ter02}). During the first, normal case, see Fig.\ \ref{fig:1direct_inderect} (b), basal ganglia network activity allows a  relatively faithful response of thalamus, closely following the sensorimotor cortex input. 
The second case considered is the parkinsonian state, depicted schematically in Fig.\  (\ref{fig:1direct_inderect}(c)), with resulting overall increase in inhibitory projections to the thalamus. 
The third case considers the situation of therapeutic intervention with deep brain stimulation (DBS) to STN, which is simulated as a high-frequency current injection into STN neurons, resulting in switching the entire network dynamics and restoring functional thalamic output close to normal behaviour. 

In order to capture the transitions and to measure qualitatively the impact of the network dynamics, we additionally defined macroscopic variables or observables such as the \emph{synchronisation index} and the \emph{average synaptic activity}, focusing on GPi as output region of the basal ganglia network. 

All simulations are made with Matlab 2020a using the solver ode23, a adaptive time step integration Runge Kutta scheme. The order of the method is three and the default relative and absolute tolerances in matlab  $[10^{-6},10^{-6}]$ are used.

\subsection{Modelling the normal state}
\begin{figure}
\begin{center}
\begin{picture}(300,300)
%\put(0,0)
\includegraphics[width=.9\textwidth]{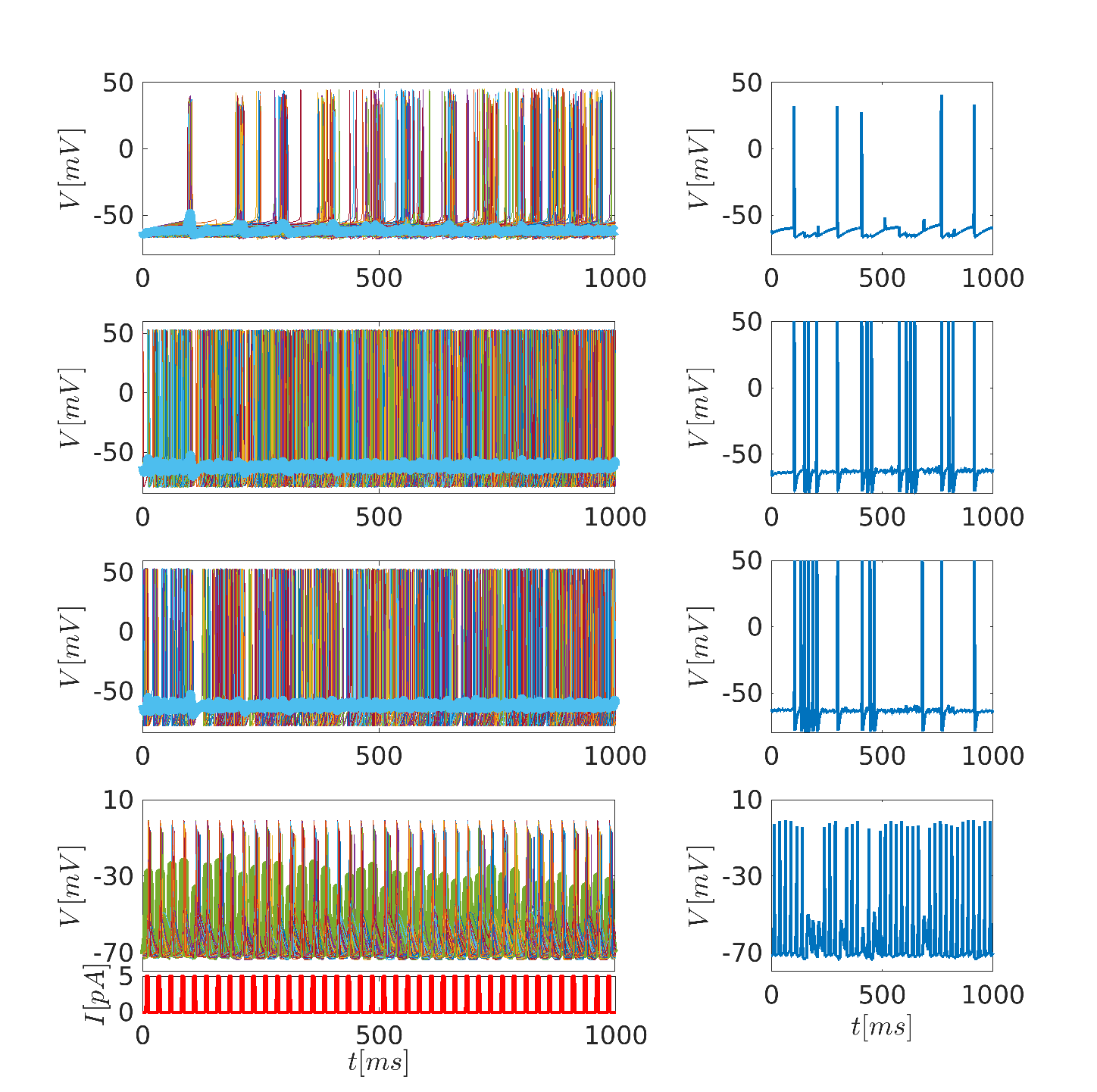}%{normalfin.eps}% This is a *.eps file
\put(-320,265){\textbf{(a)}}
\put(-320,200){\textbf{(b)}}
\put(-320,140){\textbf{(c)}}
\put(-320,75){\textbf{(d)}}
%\put(-320,20){\textbf{(e)}}
\put(-320,22){{sensori-}}
\put(-320,13){{motor input}}
\put(-20,265){STN}
\put(-20,200){GPe}
\put(-20,140){GPi}
\put(-20,75){Tha}
%\put(-20,40){ sensorimotor}
%\put(-10,30){     input}
\end{picture}
\end{center}
\caption{Time series representation of the whole network dynamics under healthy conditions (Fig.\ \ref{fig:1direct_inderect}(b)) and on the right column (insets), a time series of one representative neuron, assuming irregular STN activity resulting from a continuous current input to STN (simulating afferent synaptic input). At the same time, the thalamus receives periodic sensorimotor input (simulated by rhythmic 5 pA current injections). (a) Under these conditions, the time-dependent activity of all neurons in STN results in non-synchronised firing at approx. 4-7 Hz. (b) In GPe, as the plot of time-dependent activity of all neurons shows, this results in higher-frequency bursting not strictly related to STN activity. (c) In GPi, in turn, the time-dependent activity plot shows that GPe activity is translated to lower-frequency irregular bursting of all neurons. (d) Lastly, the time-dependent activity plot of all neurons in thalamus demonstrates that the combined input from GPi and sensorimotor drive results in a periodic firing relatively faithfully following cortical input.Insets show activity of one neuron under each condition}
\label{fig:Normal}
\end{figure}
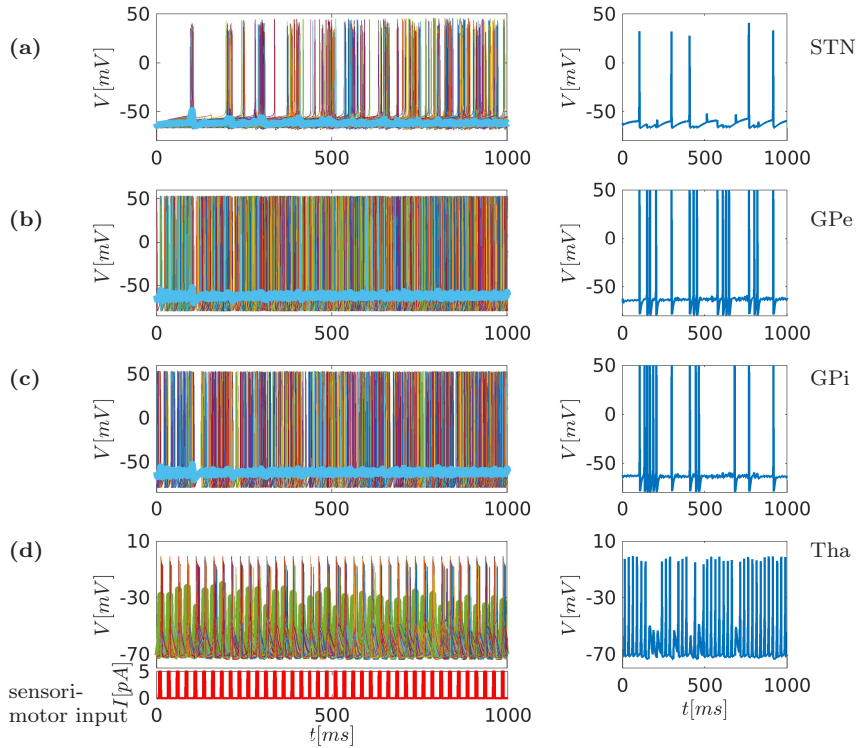

Parameters were tuned to simulate normal-healthy conditions as shown in Fig.\ \ref{fig:1direct_inderect}(b). Inhibitory signals from the striatum to the GPe are described by the current  $I_{\text{app}}=5 \text{pA}$ whereas from striatum to GPi the current has the value $I_{\text{app}}=4$pA. Fig.\ \ref{fig:Normal} depicts the time dependent activity of all nuclei under normal conditions. 
The STN neurons fire irregularly at around 4-7 Hz, i.e.\ close to values reported int he literature (cf.\ \cite{Bev99} Rubin et al 2012), see  Fig.\ \ref{fig:Normal}(a). The GPe activity is plotted in Fig.\ \ref{fig:Normal}(b). It is characterised by high-frequency irregular bursting firing with individual clustering of action potential series (which is not visible on population scale, but only on the level of a single neuron as shown in the right column). Overall, there is little correlation between STN and GPe activity. Similar dynamics appear in GPi, which in turn, also shows high-frequency firing. Looking more closely, there is, however, an underlying weak rhythmicity at approx. 12 Hz, i.e.\ in the $\beta$-band, see Fig.\ \ref{fig:Normal}, and also \ref{fig:MeanSynapGPi}(a). The resulting time-dependent activity plot of all neurons in the thalamus is shown in Fig.\ \ref{fig:Normal}(d), demonstrating that under these conditions, the thalamus faithfully responds to sensorimotor input (Fig.\ \ref{fig:Normal}(e)), without however being tightly phase-locked as in the case of single neuron, Fig.\ \ref{fig:4Thalam}. Accordingly, the thalamic response efficacy $R$ (a macroscopic variables which measures the response of the thalamic neurons to sensorimotor input, with values in the interval [0,1] and the value 1 corresponds to a whole thalamus activation, exact definition in sec.\ \ref{sec:DBS_Effic}) (see  Fig.\ \ref{fig:Effic}) in section \ref{sec:DBS_Effic}) is approximately 0.5 under normal conditions. In conclusion, the GABAergic synaptic output from the GPi to the thalamus converts thalamic responses (otherwise phase locked to the cortical input) to more loosely firing of the thalamus, without subduing it altogether as in the parkinsonian state (see Fig.\ \ref{fig:Park}).

\subsection{Modelling the parkinsonian state}

As outlined above, in Parkinson's disease, a degeneration of nigrostriatal  dopaminergic neurons, accompanied by a reduction in the number of dendritic spines of striatal medium spiny neurons \cite{Gag17, Fio16}, leads to a loss of dopamine in the striatum. The resulting overall reduction of D1/D2 receptor-mediated activity affects the direct/indirect pathway functionality. In the direct pathway, the (D1) receptors  malfunction results in disfacilitation of striatal projection neurons, and as a consequence a reduced inhibition of the GPi neurons. Thus, the disinhibited GPi increases its neuronal activity,sending higher levels of inhibition to the thalamus. 

In the indirect pathway, loss of D2-receptor activation will disinhibit striatal projection neurons. These, in turn, now decrease the activity of the GPe, to which they project. By inhibiting the GPe, the activation of STN is increased, and the overactive STN will enhance the neuronal activity in the GPi - which again leads to even more pronounced thalamic inhibition, see Fig.\ \ref{fig:1direct_inderect}(c). 

\begin{figure}
\begin{center}
\begin{picture}(300,300)
%\put(0,0)
\includegraphics[width=0.9\textwidth]{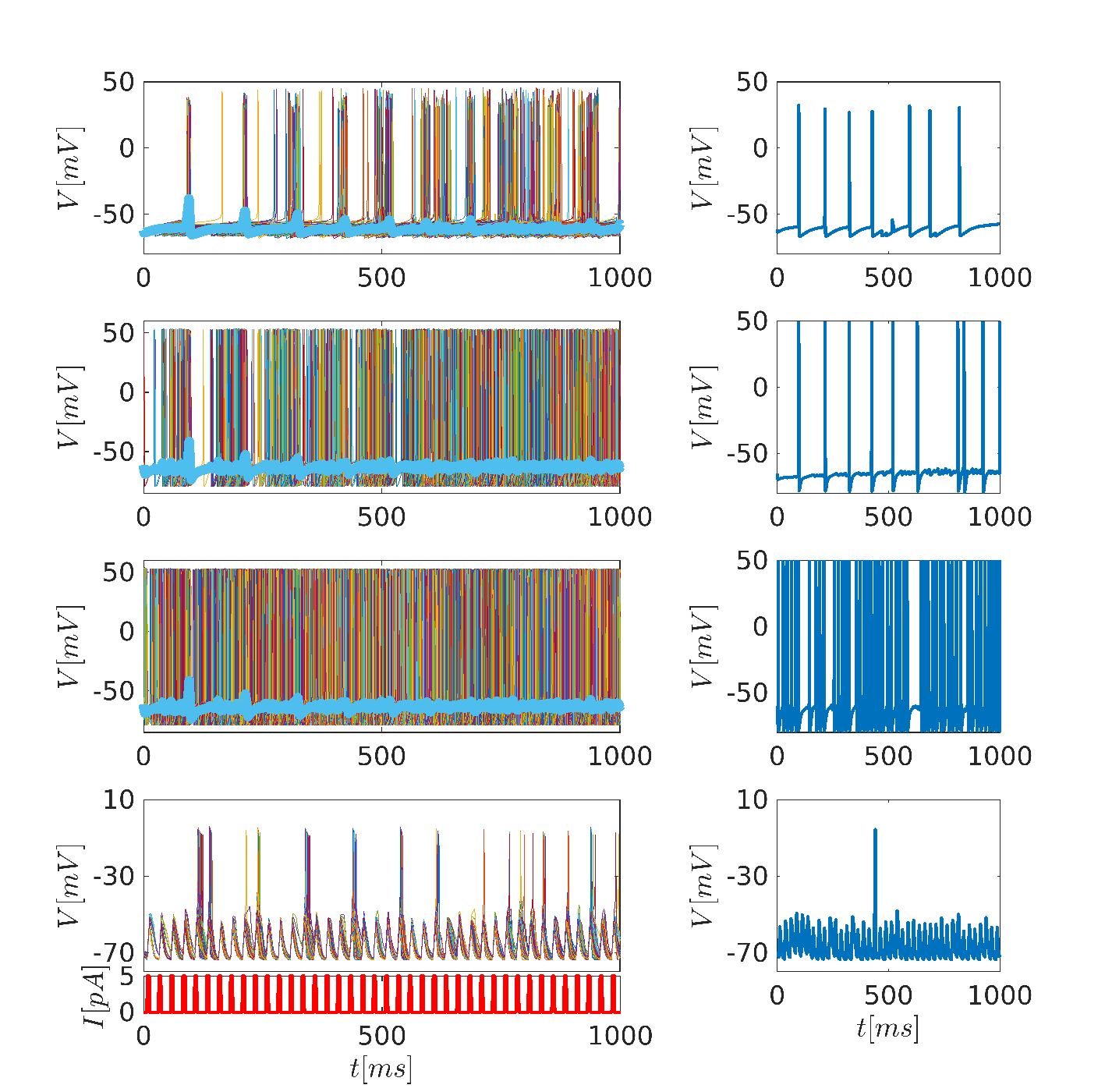}%{parkfin.eps}% This is a *.eps file
\put(-320,265){\textbf{(a)}}
\put(-320,200){\textbf{(b)}}
\put(-320,140){\textbf{(c)}}
\put(-320,75){\textbf{(d)}}
%\put(-320,30){\textbf{(e)}}
\put(-320,22){{sensori-}}
\put(-320,13){{motor input}}
\put(-20,265){STN}
\put(-20,200){GPe}
\put(-20,140){GPi}
\put(-20,75){Tha}
%\put(-20,40){ sensorimotor}
%\put(-10,30){     input}
\end{picture}
\end{center}
\caption{Time series representation of the whole network dynamics under parkinsonian conditions (Fig.\ \ref{fig:1direct_inderect}(c)), while on the right column a time series of one representative neuron (insets). We simulate synaptic input to the STN, and thalamic periodic sensorimotor input identical to the normal condition. (a) Under parkinsonian conditions, the continuous current input to STN (simulating afferent synaptic input) results in an increase in STN activity (raising the frequency from ~7 to ~11 Hz) at a more synchronised level (see occasional simultaneous firing of one cell in the right column) (b) In GPe, as the plot of time-dependent activity of all neurons shows, this results in a reduction of firing frequency compared to normal conditions, i.e. low frequency bursting which is more coupled to STN activity (compare in right column). (c) In GPi, in turn, the time-dependent activity plot shows that this altered GPe activity is translated to bursting at a frequency 4-5 times higher than under normal conditions, which should result in strong inhibition of the thalamus. (d) Indeed, this inhibition is reflected in the time-dependent activity plot of all neurons in thalamus. It demonstrates that the combined input from GPi and sensorimotor drive now results in very sparse firing, and essentially a loss of sensorimotor-thalamic coupling.}
\label{fig:Park}            
\end{figure}
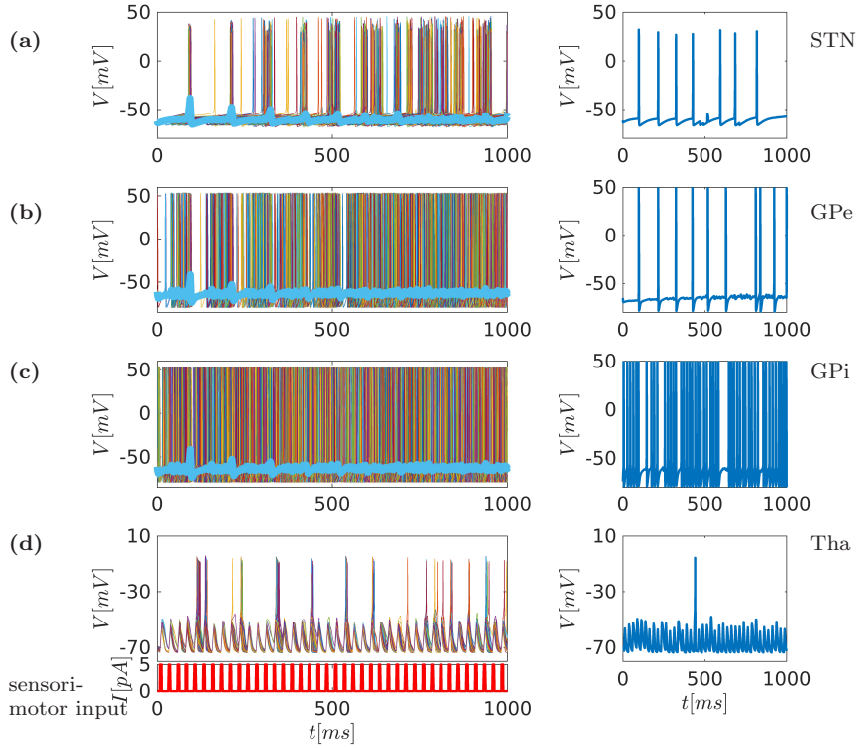

Consistent with this concept, the model assumes a decrease in the level of inhibition of GPi neurons $I_{\text{app}}$. This is modelled by increasing depolarising current from 4 to 8 pA (corresponding to dis-inhibition). At the same time, the model assumes an increase in the level of inhibition of GPe neurons (therefore, depolarising current is decreased from 5 to 3pA). Fig.\ \ref{fig:Park} shows the dynamics of the network in time of the four areas:
Under parkinsonian conditions, the time-dependent activity of all neurons in STN is changed compared to the normal state: Firing becomes more regular, and occurs at higher frequency (approx. 11 Hz). For the GPe Fig.\ \ref{fig:Park}(b), the consequence is a lower burst frequency, almost following the STN activation (see insets). Fig.\ \ref{fig:Park}(c) shows that in GPi, the altered activity is translated to high-frequency bursting, with an underlying enhanced rhythmicity in $\beta$-activity (13-15 Hz) compared to the normal condition, see also Fig.\ \ref{fig:MeanSynapGPi}(b).  As a result, inhibition to the thalamus strongly increases, and the thalamus is no longer able to transmit signals in response to sensorimotor input, as its firing becomes very sparse.

\subsection{Modelling the effects of DBS}

\begin{figure}
\begin{center}
\begin{picture}(300,310)
\includegraphics[width=.9\textwidth]{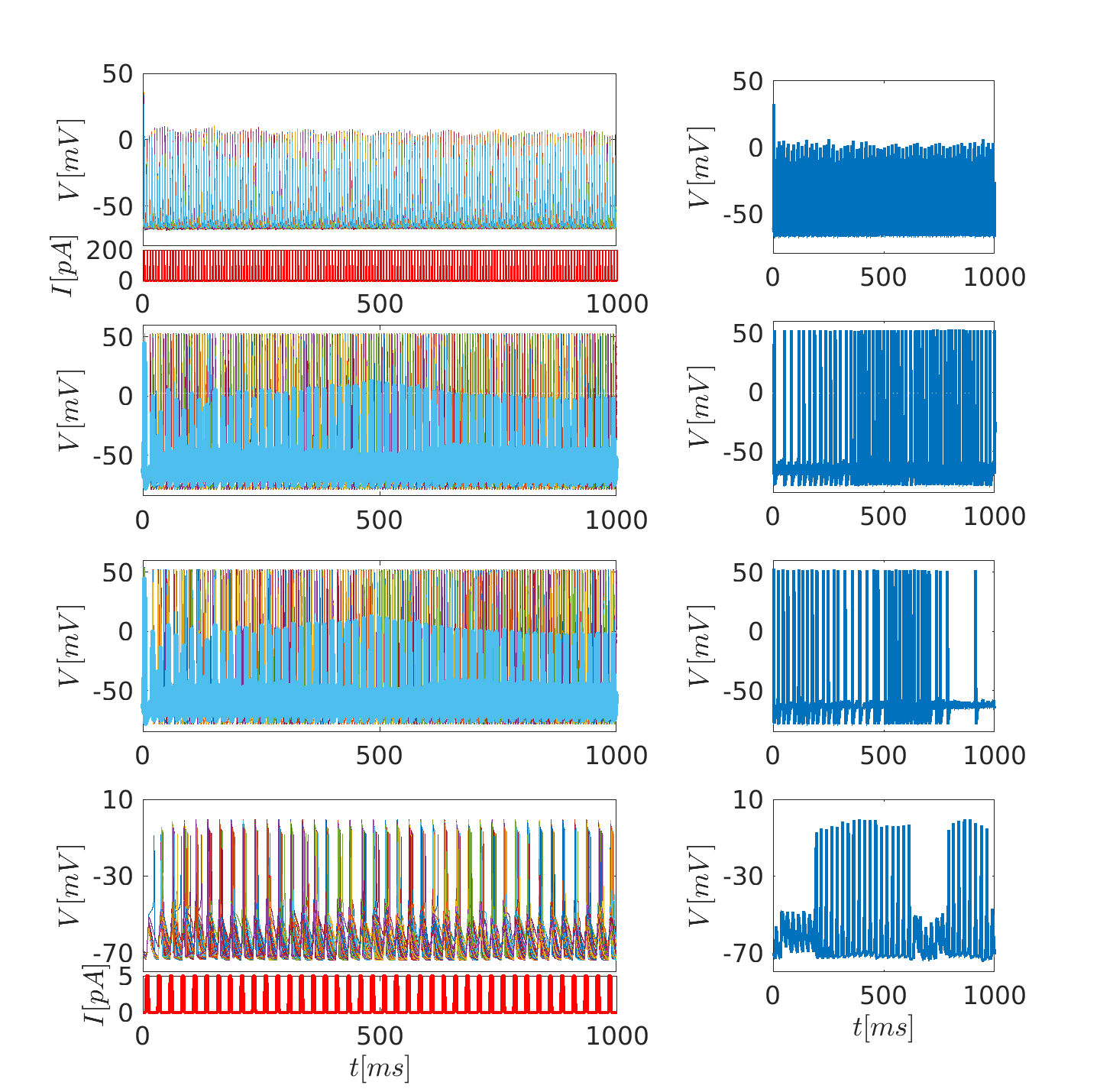}%{DBSparkfin.eps}% This is a *.eps file
\put(-320,265){\textbf{(a)}}
\put(-315,220){DBS}
\put(-320,200){\textbf{(b)}}
\put(-320,140){\textbf{(c)}}
\put(-320,75){\textbf{(d)}}
%\put(-320,30){\textbf{(e)}}
\put(-320,22){{sensori-}}
\put(-320,13){{motor input}}
\put(-20,265){STN}
\put(-20,200){GPe}
\put(-20,140){GPi}
\put(-20,75){Tha}
%\put(-20,40){ sensorimotor}
%\put(-10,30){     input}
%\put(14.7,11.7){STN}
%\put(14.1,11.){DBS input}
%\put(14.7,9.1){GPe}
%\put(14.7,6.5){GPi}
%\put(14.7,3.9){Tha}
%\put(13.7,1.8){ sensorimotor}
%\put(14.4,1.2){  input}
\end{picture}
\end{center}
\caption{Time series representation of the whole network dynamics under parkinsonian conditions (Fig.\ \ref{fig:1direct_inderect}(c), but now simulating DBS by high-frequency current injection into STN. At the same time, the thalamus continues to receive periodic sensorimotor input (simulated by rhythmic 5 pA current injections). (a) Under parkinsonian conditions and DBS simulation, the time-dependent activity of all neurons in STN converts from periodic into tonic high-frequency firing, following the DBS input. (b) In GPe, as the plot of time-dependent activity of all neurons shows, this results similarly tonic, regular firing at a slightly lower frequency (elevated compared to normal and even more so compared to parkinsonian state). (c) In GPi, in turn, the time-dependent activity plot shows that this altered GPe activity is translated to slightly irregular, constant firing at even lower frequency, very close to the normal state. Losing the high-frequency discharges in GPi , the inhibitory drive to the thalamus is reduced, which is reflected also in the reduction of the mean activity index, see also Fig.\ \ref{fig:MeanSynapGPi}(c).   
(d) This disinhibition is reflected in the time-dependent activity plot of all neurons in thalamus. It demonstrates that the combined input from GPi and sensorimotor drive, together with DBS to STN, now stably restores firing in approximate synchrony to sensorimotor input.}
%{BS-treatment dynamics. DBS activity at 130HZ (in current pulsing form) is applied on STN. The inhibition levels of striatum was kept the same as the Parkinson. The network dynamics change dramatically from the Parkinsonian case (Fig. \ref{fig:Park}).  The STN fires continuous with almost zero refractory period(first row). This overactive STN response to DBS is relocated according to network topology to the other nuclei. Both GP (internal-external, seond and third row) become also overactive but new synchronised pattern.   with enhanced  bursting activity. As result, a strong inhibition signal is transferred to 0,
%thalamus which now depicts abnormal behaviour. The inefficient  reduced neuronal activity preventing thalamus to respond faithful to the external PFC stimulus. The GP activity can be characterised as strong beta bursting activity similar to cite[Plenz and Kital nature].}
\label{fig:DPS_T}
\end{figure}
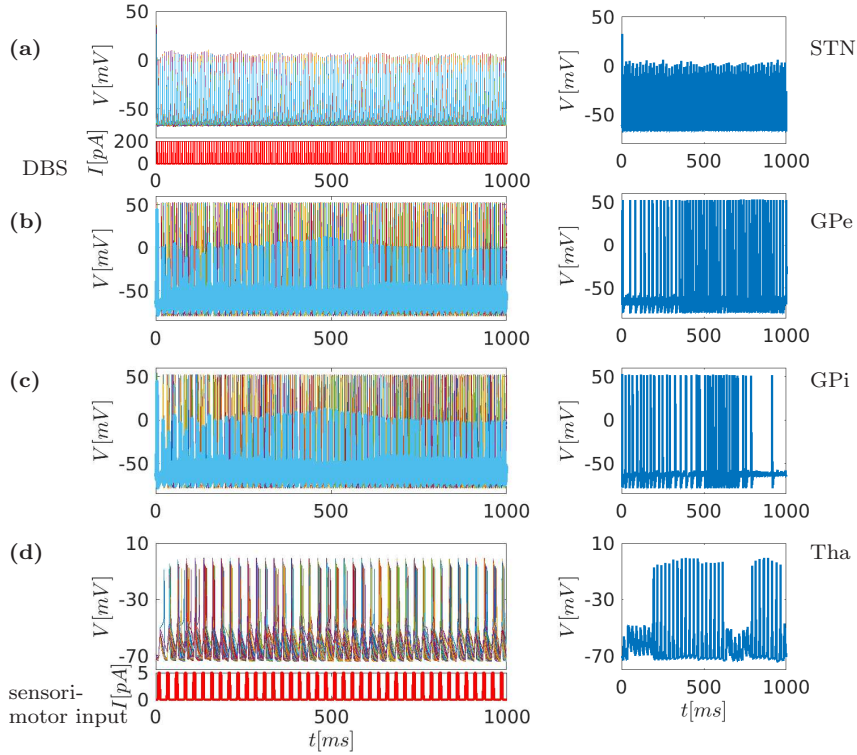

The network structure and the conductances were kept invariant with respect to parkinsonian conditions. DBS treatment was simulated by a high frequency current of 184 Hz, applied to all STN neurons (this value resulted after the analysis in section \ref{sec:DBS_Effic} and is suggested as candidate for an optimal DBS frequency in experiments). The high frequency current is modelled as periodic short pulses of the form \cite{Rub04}

\begin{equation}
    I_{\text{DBS}}=A_{\text{DBS}}H(\sin(2\pi t/T_{\text{DBS}})\cdot (1-H(\sin(2\pi(t+\delta_{\text{DBS}})/T_{\text{DBS}})) .
\end{equation}
In all three basal ganglia nuclei, DBS induces dramatic alternations in firing dynamics. Regarding the STN, neurons follow the strong DBS signal and fire tonically at stimulation frequency, see Fig.\ \ref{fig:DPS_T}(a). In GPe, this results in regular bursting of neurons at about 90 Hz, i.e. at ~3x higher frequency than normal, and ~4x higher frequency than under parkinsonian conditions, respectively. In GPi, in turn, the increased activity in GPe nearly normalises the firing to slightly irregular firing at ~30 Hz (compare insets in this figure and \ref{fig:Normal}), and thus strongly reduces firing frequency compared to the parkinsonian state (which ranged around 50-80 Hz) see Fig.\ \ref{fig:DPS_T}. 
In summary, there is an overall loss of synchronisation between STN, GPE and GPi, and at the same time STN and GPe now fire tonically. With the reduction of the high-frequency tonic firing of the GPi, the thalamus is disinhibited and resumes firing, following the sensorimotor input relatively closely. 
 
\subsection{STN-GPi interaction: synchronisation changes and local travelling waves formation}

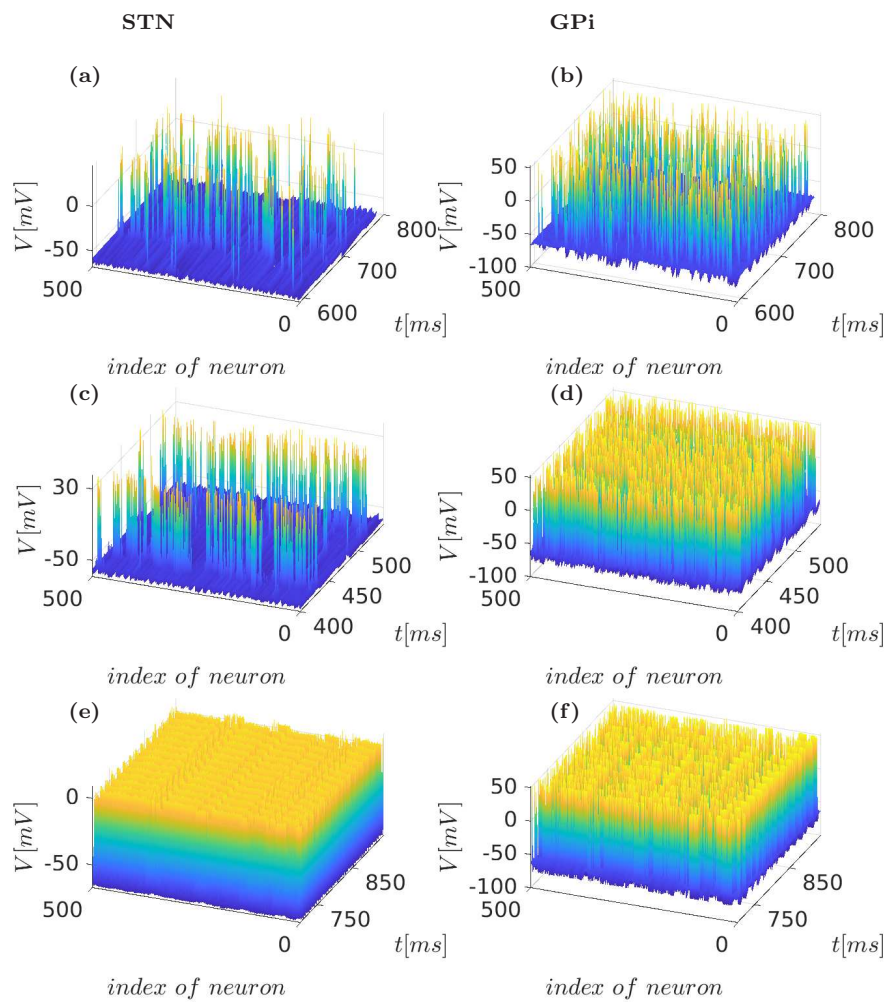
\begin{figure}
\begin{center}
\begin{picture}(330,400)
%\put(0,0)
\includegraphics[width=1.1\textwidth]{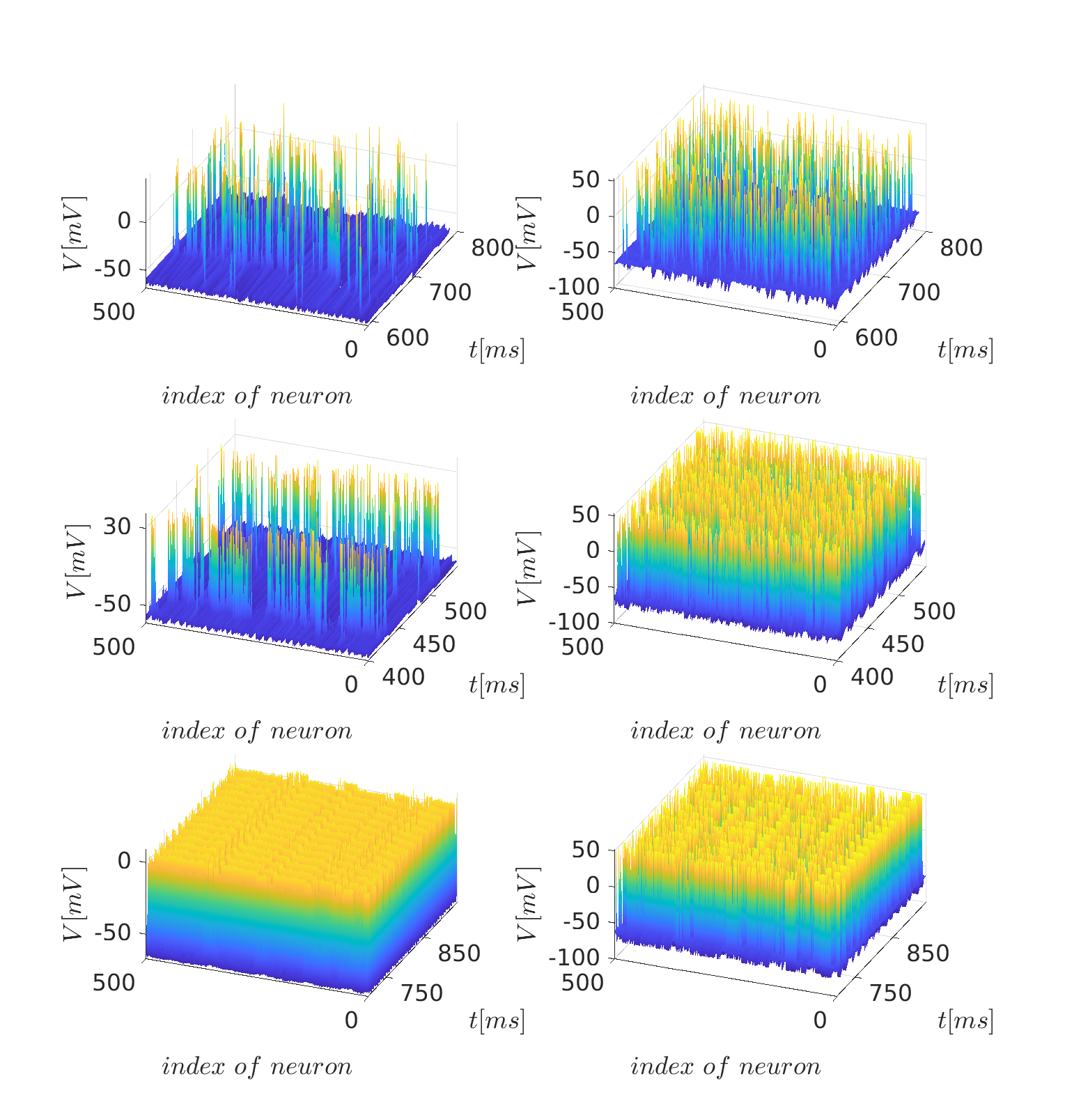}% This is a *.eps file
\put(-310,380){\textbf{STN}}
\put(-150,380){\textbf{GPi}}
\put(-330,360){\textbf{(a)}}
\put(-150,360){\textbf{(b)}}
\put(-330,240){\textbf{(c)}}
\put(-150,240){\textbf{(d)}}
\put(-330,120){\textbf{(e)}}
\put(-150,120){\textbf{(f)}}
%\put(-330,100){\textbf{(g)}}
%\put(-150,100){\textbf{(h)}}
\end{picture}
\end{center}
\caption{Spatio-temporal activity patterns for STN (left column) and GPi (right column) under normal (a and b), parkinsonian (c and d) and DBS (e and f) conditions. Colour code depicts voltage (i.e. activity level) against time (in ms) and space (i.e. index of neuron of the nuclei see also Fig.\ \ref{fig:5net}.   
}
\label{fig:travel_welle}            
\end{figure}

\begin{figure}
\begin{center}
\begin{picture}(330,230)

\includegraphics[width=1.\textwidth]{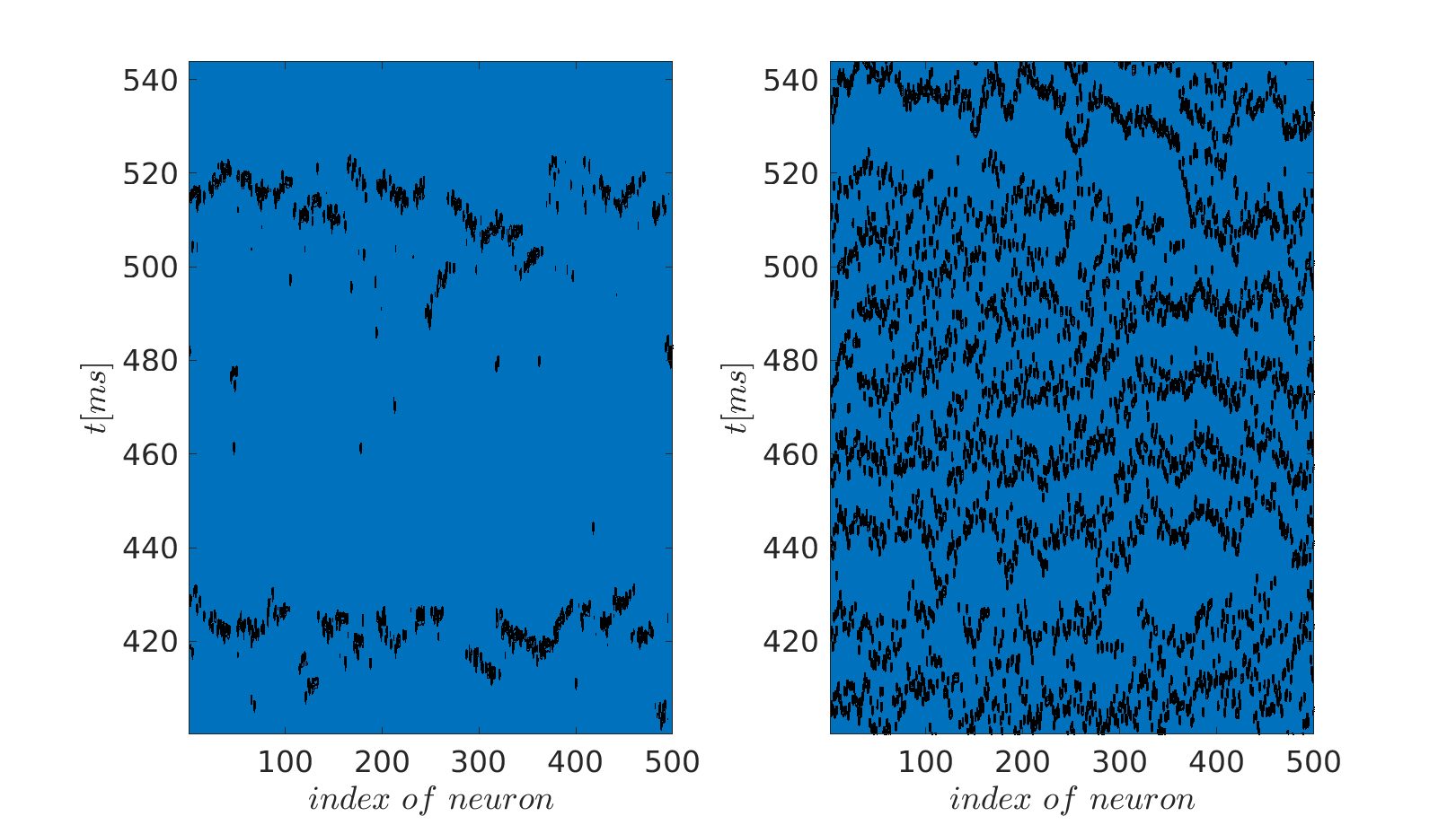}% This is a *.eps file
%\put(-330,210){\textbf{(a)}}
%\put(-150,210){\textbf{(b)}}
%\put(-310,230){\textbf{STN}}
%\put(-150,230){\textbf{GPi}}
\put(-295,210){\textbf{(a)}}
\put(-150,210){\textbf{(b)}}
\put(-275,210){\textbf{STN}}
\put(-130,210){\textbf{GPi}}
%\put(-330,100){\textbf{(g)}}
%\put(-150,100){\textbf{(h)}}
\end{picture}
\end{center}
\caption{Spatio-temporal activity patterns for STN (a) and GPi(b) in parkinsonian state. Both areas shows slow travelling waves substructures, with stronger indication in GPi area. This dynamics coexisting with synchronous (macroscopic) activity. 
}
\label{fig:travel_welle3}            
\end{figure}
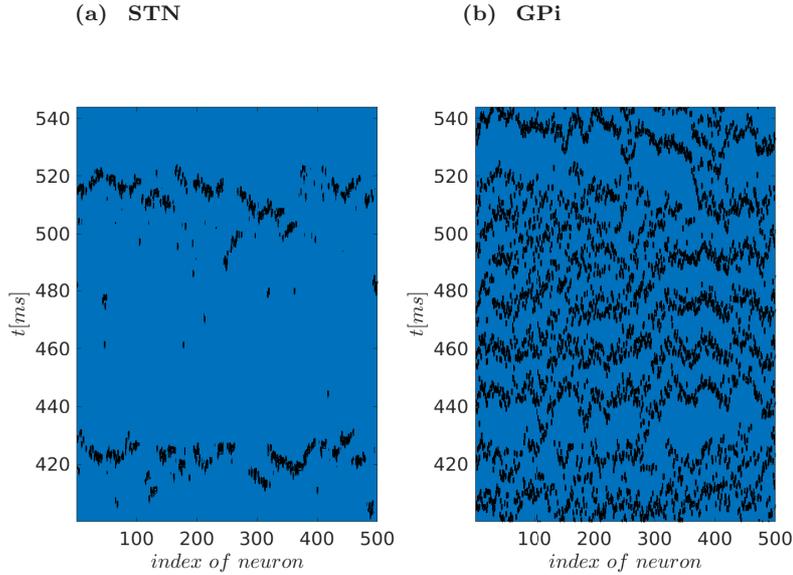

Different spatio-temporal patterns of activation in STN on the one hand, and GPi on the other, can be observed under normal, parkinsonian and DBS conditions. Fig.\ \ref{fig:travel_welle} depicts the whole neuronal activity of STN and GPi regions in the case of to weak-sparse  connectivity between STN neurons. In normal case (a) and (b), both ares STN and GPi show irregular patterns. Instead, in parkinsonian case (c) and(d), the regions show higher synchronous activation. This is also confirmed in Fig.\ \ref{fig:travel_welle3}: This raster plot shows the sparse, but very synchronous firing in STN in the parkinsonian case, with local travelling waves, restricted to a few neurons and demonstrating some clustering of neuronal activity. A reasonable interpretation for this is that in the parkinsonian state, the reduced inhibition from GPe to the STN leads to a higher spatial synchronisation between these nuclei, resulting in increased order in the pattern formation.

Which consequence does this have for the activity in the GPi?

This nucleus receives both input from the STN (indirect pathway) and from the striatum (direct pathway), and hence one would expect two competing activity patterns. Under parkinsonian conditions, in turn, the inhibitory control of the direct pathway is reduced. Hence the activity in the GPi is more pronounced, as also confirmed in Fig.\ \ref{fig:travel_welle3}, which shows series of quasi synchronous activations of GPi neurons, with local travelling waves forming small clusters. 
Under DBS conditions, the massive synchronisation in the STN (abolishing all travelling waves and imposing a 184 Hz rhythm on nearly all neurons) is mirrored in a closely-matching synchronised activity pattern in the GPi, but at lower frequency of around 60Hz, again overriding all other activity. This presumably results in a nearly tonic activation of inhibitory neurons, to which the thalamic neurons desensitise, as will be discussed below.

 \subsection{Effects of dense STN connectivity}
\label{sec:dense}
\begin{figure}
\begin{center}
\begin{picture}(330,430)
%\put(0,0)
\includegraphics[width=1\textwidth]{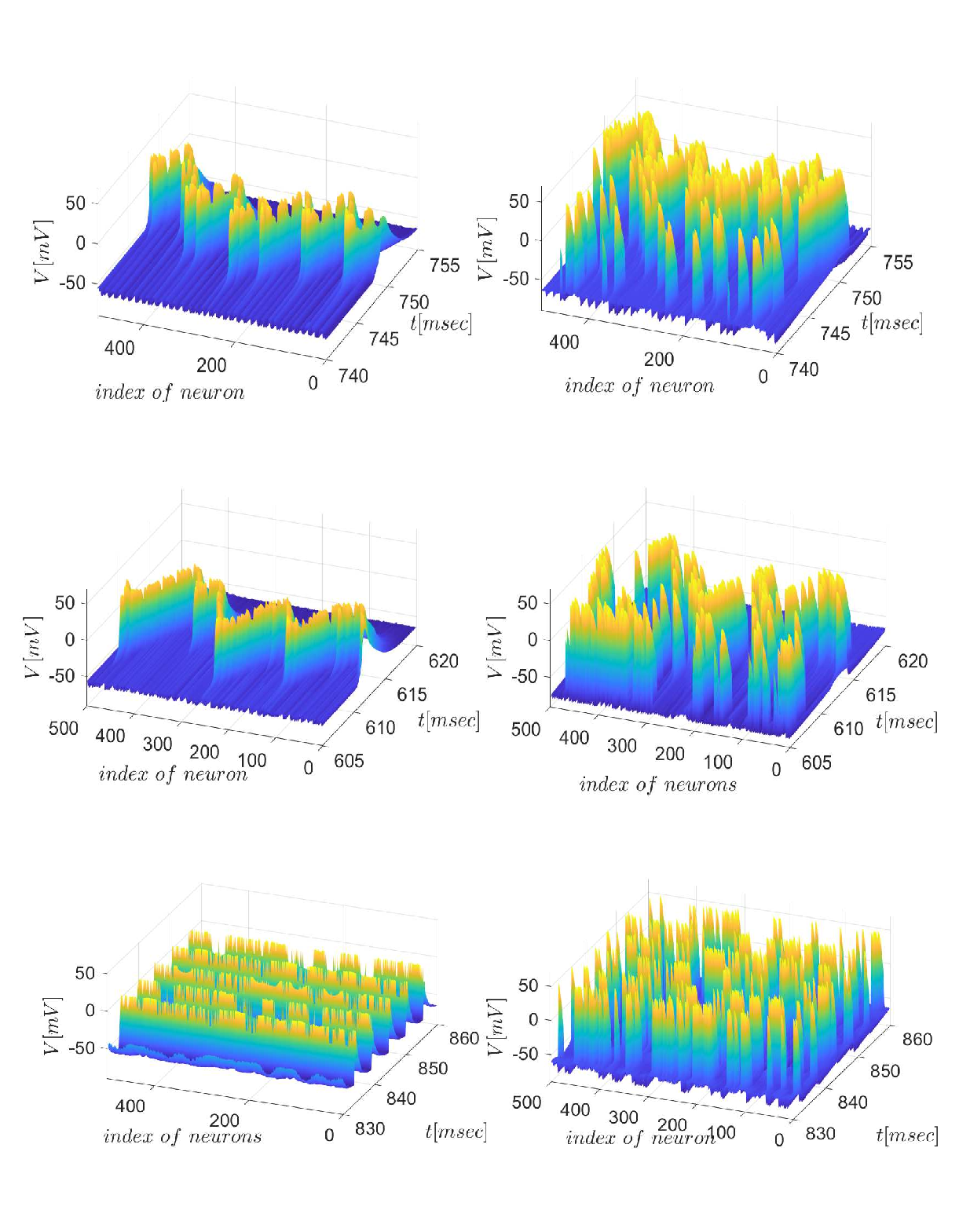}% This is a *.eps file
\put(-310,400){\textbf{STN}}
\put(-150,400){\textbf{GPi}}
\put(-330,380){\textbf{(a)}}
\put(-150,380){\textbf{(b)}}
\put(-330,240){\textbf{(c)}}
\put(-150,240){\textbf{(d)}}
\put(-330,120){\textbf{(e)}}
\put(-150,120){\textbf{(f)}}
%\put(-330,100){\textbf{(g)}}
%\put(-150,100){\textbf{(h)}}
\end{picture}
\end{center}
\caption{Spatio-temporal activity patterns of basal ganglia network with the assumption of strong connectivity of STN area, for STN (left column) and GPi (right column) under normal (a and b), parkinsonian (b and c) and DBS (e and f) conditions. Colour code depicts voltage (i.e. activity level) against time (in ms) and space (i.e. index of neuron of the nuclei see also Fig.\ \ref{fig:5net}.The spatio-temporal activity in the regions of STN and GPi strong indicates travelling wave solutions with more variable structures in the normal state.
}
\label{fig:travel_welle2}            
\end{figure}
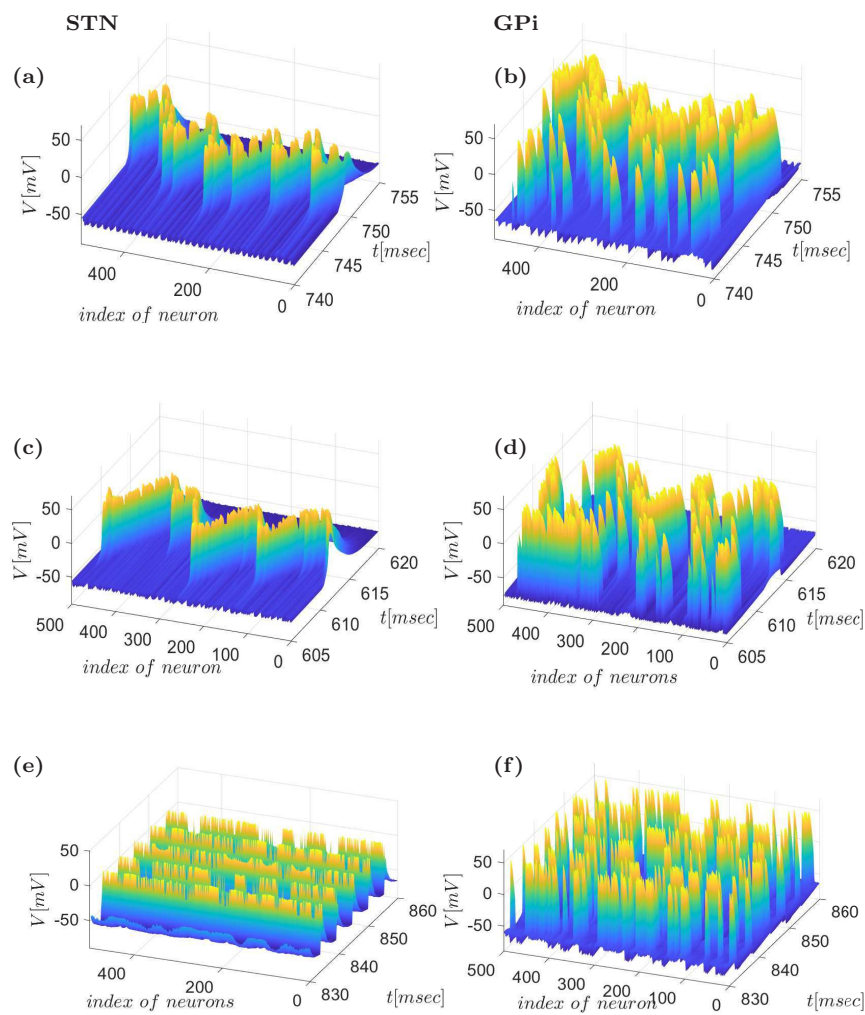

 Since the degree of conncetivity among STN neurons is still under debate \cite{Stein19}, in a next step also a higher connectivity was considered. The increment of connectivity, i.e., higher number of connections between STN neurons, leads to different spatio-tempral pattern. Fig. \ \ref{fig:travel_welle2} shows the normal, parkinsonian and DBS states. The qualitatively difference here is the strong connectivity within  STN. Each neuron now has mutual connections with other neurons with mean number of connections equal to 20. The STN structure  is described  as a small world  topology. Under normal and parkinsonian conditions, the basal ganglia network depicts strongly correlating travelling waves, which propagate, and are more variable in the normal state. The travelling wave represents a propagation of similar activity levels along one direction, i.e.\ as activity peaks in neighbouring neurons in a co-moving frame, preserving the activity shape. Also with this much higher intra-STN connectivity, the parkinsonian state is characterised by increased synchronisation in GPi. Under DBS, rhythmic activity of STN in turn is reflected in less synchronisation in GPI. Thus, the model is rather robust with respect to GPi output, within a wide range of STN connectivity. 

\section{Synchronisation and synaptic activity indices characterise transitions of network dynamics}
In order to shed light on the transitions between the different states, and to examine the appearance of distinct patterns in our model, we analysed both the level of synchronisation within the GPi (computing the synchronisation index $r$), and the levels of synaptic GABAergic projection activity from the GPi to the thalamus (defining GPi activity index $l$). 

The theory of phase synchronisation \cite{Izhi04,Erm12,Tass99,Piko01,Kur84} described in section \ref{sec:synch}, allows us to characterise and analyse different attributes and the dynamics which results from the mathematical model. The main observables are the synchronisation index $r$ defined in eq.\ \eqref{eq:Sync_index} and the mean GPi synaptic activity index $l$ defined in eq.\ \eqref{eq:LFP} and depicted in Figs.\ \ref{fig:Synchidex_allcases} and  \ref{fig:MeanSynapGPi}. We further performed a Fourier analysis of the mean GPi activity index, in order to measure the dominant frequencies (rhythms) and we determined its slope of the power spectrum to estimate the distribution of frequencies in the power spectrum (see Fig.\ \ref{fig:Fouriertot}).

\subsection{Macroscopic dynamics in the normal state}

\begin{figure}
\begin{center}
\begin{picture}(330,200)
\put(10,180){\textbf{(a)}}
\put(130,180){\textbf{(b)}}
\put(230,180){\textbf{(c)}}
%\put(0,0)
\includegraphics[width=1.\textwidth]{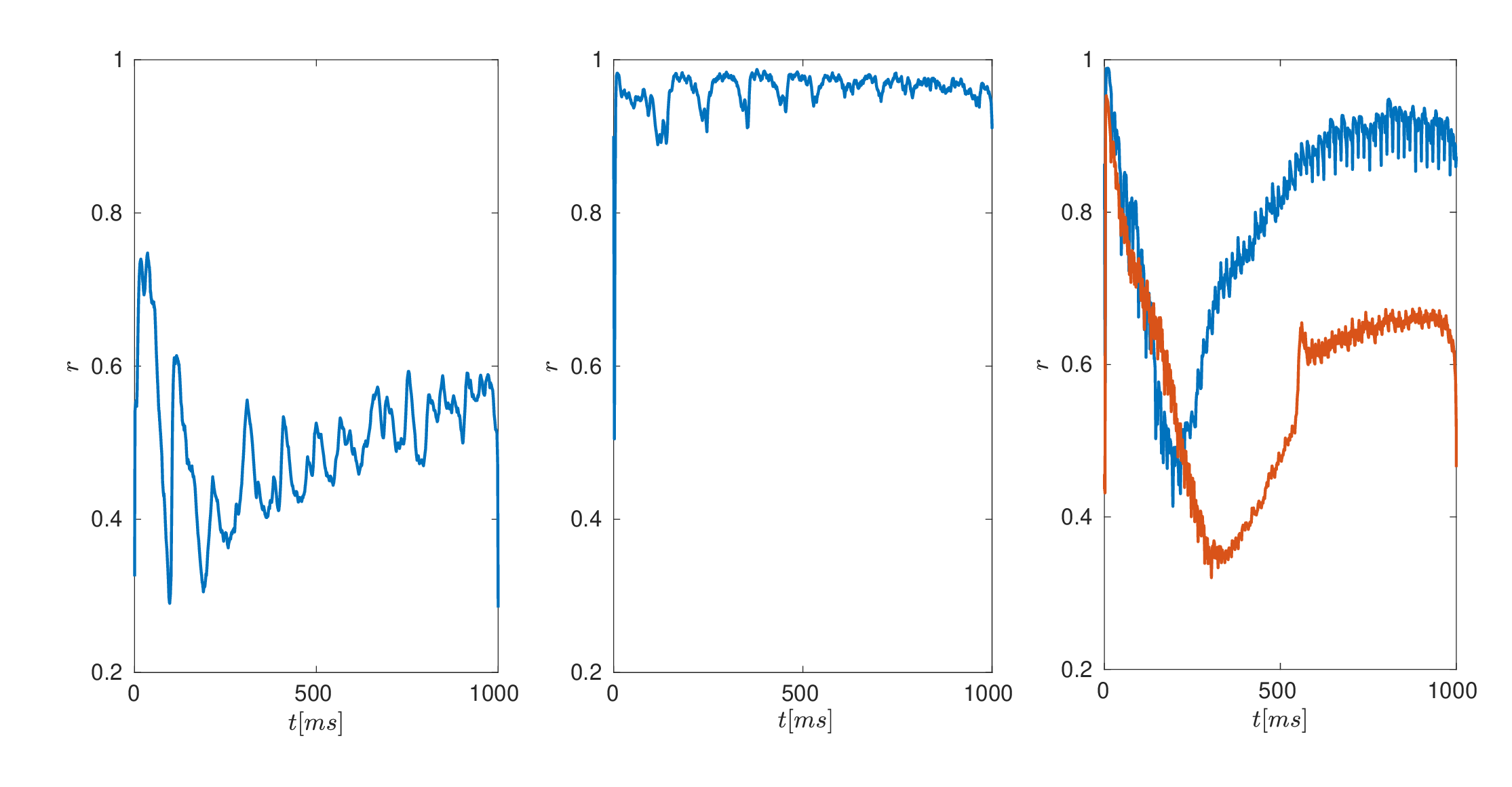}% This is a *.eps file
\end{picture}
\end{center}
\caption{Synchronisation index $r$, defined in eqs.\ (\ref{eq:Sync_index}, \ref{eq:Hilbert}) for (a) healthy normal, (b) parkinsonian state and (c) DBS. (a) In the healthy state, the synchronisation index shows an oscillatory behaviour at frequency of $\approx$11 Hz with relatively low values of range [0.3, 0.6]. (b) In the parkinsonian state, in turn, the synchronisation index is high fluctuates irregularly for longer periods during the bursting activity in the GPi. This bursting activity is highly frequent and prolonged, shifting to enhanced $\beta$ power. As a result, the inhibitory drive to the thalamus is strongly increased. 
(c)
% For 184 Hz:
Simulating STN-DBS, the synchronisation index for 184Hz and 210Hz (red curve). With both frequencies, $r$ starts out at almost 1, and then wanes down to $\approx$0.4 in periodic, high-frequency dips, with ongoing tonic firing in GPi, where $\beta$ activity is lost. Importantly the higher frequency i.e. 210Hz provides higher de-synchronisation in GPi activity. This is also coincides with decreased activity of GPi area, see Fig.\ \ref{fig:MeanSynapGPi}(c).
% For 166 Hz: Simulating STN-DBS, the synchronisation index falls to very low levels within 200 ms into the DBS stimulation, with ongoing tonic firing in the GPi, where $\beta$ activity is lost. 
}
\label{fig:Synchidex_allcases}            
\end{figure}

Considering the healthy, normal state, as Fig.\ \ref{fig:Synchidex_allcases}(a) shows, synchronisation 
within the GPi is generally low (around 0.5). The synchronisation index shows oscillatory behaviour (at around 12 Hz).  The local maxima of the synchronisation index thus correspond to increased synchronous activation of GPI neurons in the low $\beta$ range. The GPi activity results in an overall periodic, but sparse inhibitory drive to the thalamus. 
Corresponding to this behaviour, the mean synaptic GPi activity phasically oscillates around 0.2 (even if the microscopic behaviour of neurons form bursting clusters) see, Fig.\ \ref{fig:MeanSynapGPi}(a). This  rhythm lies within the lower frequencies (i.e. $\alpha$ and lower $\beta$ band), as the spectrum analysis reveals, see Fig.\ \ref{fig:Fouriertot}(a) with a first peak at 7 Hz and a higher peak at around 11 Hz. This rhythmic peak is centered relatively narrowly around this dominating frequencies, since the slope of linear approximation is relatively steep (around $-3.6$ Hz/sec), meaning that the higher frequencies do not contribute in the magnitude of the power spectrum.

\subsection{Macroscopic dynamics in the parkinsonian state}
Under parkinsonian conditions, the GPi neurons are generally much more active than in the normal state. Thus, the synchronisation is very high (close to 1). This index reflects prolonged intervals of high activity, interrupted only briefly by small decrements in synchronisation levels.
This indicates that neurons are synchronised during the prolonged and accentuated bursting in GPi in the parkinsonian state. As a result, the synaptic, inhibitory projection to the thalamus remains periodic, but predominantly strong, and is again only slightly reduced during short burst intervals, see Fig.\ \ref{fig:MeanSynapGPi}(b). The mean GPi synaptic activity (fluctuating around 0.7) is commensurate with this high synchronisation, which is supported by prolonged bursting of GPi neurons, only periodically interrupted by brief periods of reduced activity. The Fourier analysis reveals a dominant $\alpha$ /low$\beta$ activity with a strong peak at 10Hz, and a secondary at 18Hz, i.e., higher $\beta$ rhythm than the normal conditions. Importantly, the frequency spectrum of this bursting is much broader, as indicated by the corresponding linear approximation the slope increasing to $-2.5$ Hz/sec. This reflects also higher frequencies contributing to the total power.

\subsection{Macroscopic dynamics during DBS}
\begin{figure}
\begin{center}
\begin{picture}(330,200)
\put(10,180){\textbf{(a)}}
\put(130,180){\textbf{(b)}}
\put(230,180){\textbf{(c)}}
%\begin{picture}(17,8)
%\put(0,7){(a)}
%\put(6,7){(b)}
%\put(12,7){(c)}
%\put(0,0)
\includegraphics[width=1.\textwidth]{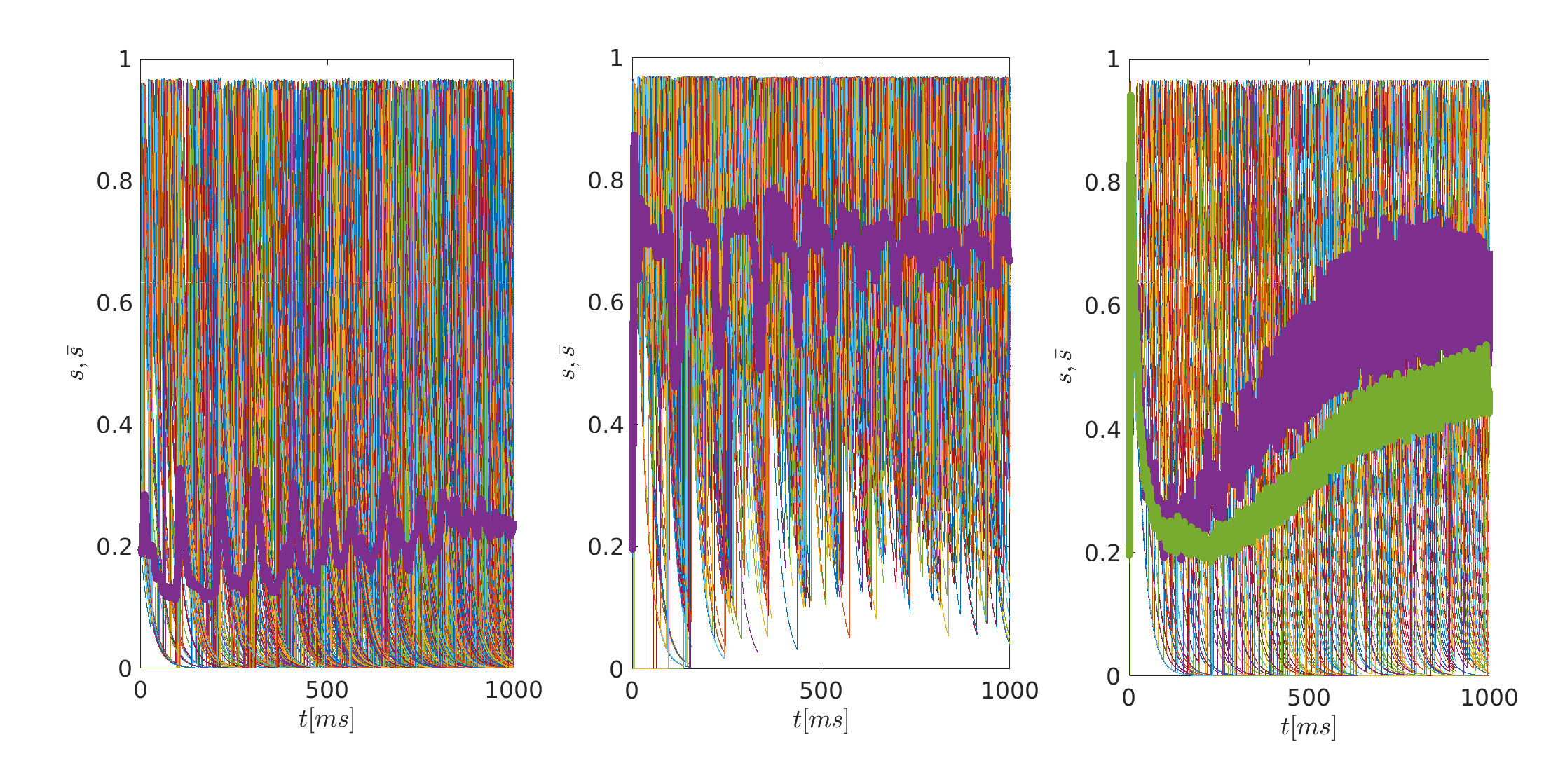}% This is a *.eps file
%\end{picture}
%\includegraphics[width=17cm]{synaptictot.eps}% This is a *.eps file
\end{picture}
\end{center}
\caption{Synaptic activity $l$ of GPi projecting to the thalamus over time $t$ defined in \eqref{eq:LFP}. Both the activities of all 500 cells (thin lines, different colours) and the mean of the activity (purple thick line) are depicted. 
%the ordinate shows probability of activation of thalamic cells
(a) Under healthy conditions, the inhibitory input to the thalamus occurs periodically in the low $\beta$ range, and more importantly, in between these inhibitory periods, the degree of synaptic GABAergic activity projecting to the thalamus is low (at $\approx$0.2). (b) Under parkinsonian conditions, in turn, the situation completely reverses: Presumably due to prolonged burst firing, the inhibitory tone is generally high ($\approx$0.7), and only intermittently during pauses of the burst firing in GPi it drops. The GPi neural activity is characterised by $\beta$ range. (c) 
Under STN-DBS for 184 and 210 Hz (light green), the inhibitory drive tonically equilibrates at a mean level of $\approx$ 0.6 for 184Hz; and around 0.5 at 210Hz. In the case of 210Hz the inhibitory tones is lower than the 184 (purple colour). %The higher frequency of 210Hz provides less inhibitory tone, as basal ganglia output to thalamus.
% 166 Hz: Under STN-DBS, the inhibitory drive tonically equilibrates at a level of 0.77; i.e. the rhythmic nature of GABAergic transmission is lost. Also the mean activity is depicted.
}
\label{fig:MeanSynapGPi}            
\end{figure}

Simulating DBS, the macroscopic activities (as the indices $r$ and $l$ show) change dramatically (depending on a specific range of frequencies), introducing a state dissimilar to both normal or parkinsonian states. The synchronisation index within the GPi shows a dynamic development over time, starting at values of $\approx$ 0.9, and then decreasing to $\approx$ 0.5, during the first 500 ms of tonic irregular firing in the GPi. After this period, the system adapts and oscillates at high frequencies between values 0.8 and close to 1, i.e.\ it remains in a relatively high synchronisation state. 
Similar dynamics emerge regarding the mean GPi synaptic activity, with a first transient period of 500 ms where the synaptic drive is decreased (to $\approx$ 0.2), and a second period ($t>500$ ms) where high synaptic activity is maintained (with fluctuations around a mean of $\approx$ 0.6, i.e. higher than under normal conditions, and slightly lower than under parkinsonian). Surprisingly, this would appear to impose a strong inhibitory drive to the thalamus, see Fig.\ \ref{fig:MeanSynapGPi}(c). 
This, however, differs from the parkinsonian condition as it is tonic high-frequency, and not repeating bursting blocks appearing at $\beta$ frequency. This qualitative change in rhythmicity is crucial: While a relatively high inhibitory tonic drive seems to suggest strong thalamic inactivation, the contrary is the case. Due to the fact that the input to the thalamus is not rhythmic, but high-frequency tonic (peak power at 184 Hz), the GABAergic synapses are deactivated (see eqs.\ \eqref{eq:synaptic_Gen}. Indeed, the frequency analysis of the GPi activity shows a main peak seeming to resonate with the external DBS stimulus (around 60Hz), and corresponding harmonics. Importantly, the peaks in the lower (normal state) and higher (parkinsonian state) $\beta$ band disappear, while the slope changes back to a value around -3.1 Hz, close to the normal case. Although this slope is close to normal conditions, it now reflects a narrow high-frequency band, and not unclustered activity. 

Taken together, the findings underscore the importance of synchronous regular and brief, clustered burst firing in the GPi for successful inhibition in the thalamus, which normally takes place at $\beta$ frequency. Although DBS thus does not restore normal rhythmicity in the GPi, the thalamic activity is disinhibited by the loss of synchronisation and clustering within the GPi. 

\begin{figure}
\begin{center}
\begin{picture}(330,200)
\put(10,180){\textbf{(a)}}
\put(130,180){\textbf{(b)}}
\put(230,180){\textbf{(c)}}
\includegraphics[width=1\textwidth]{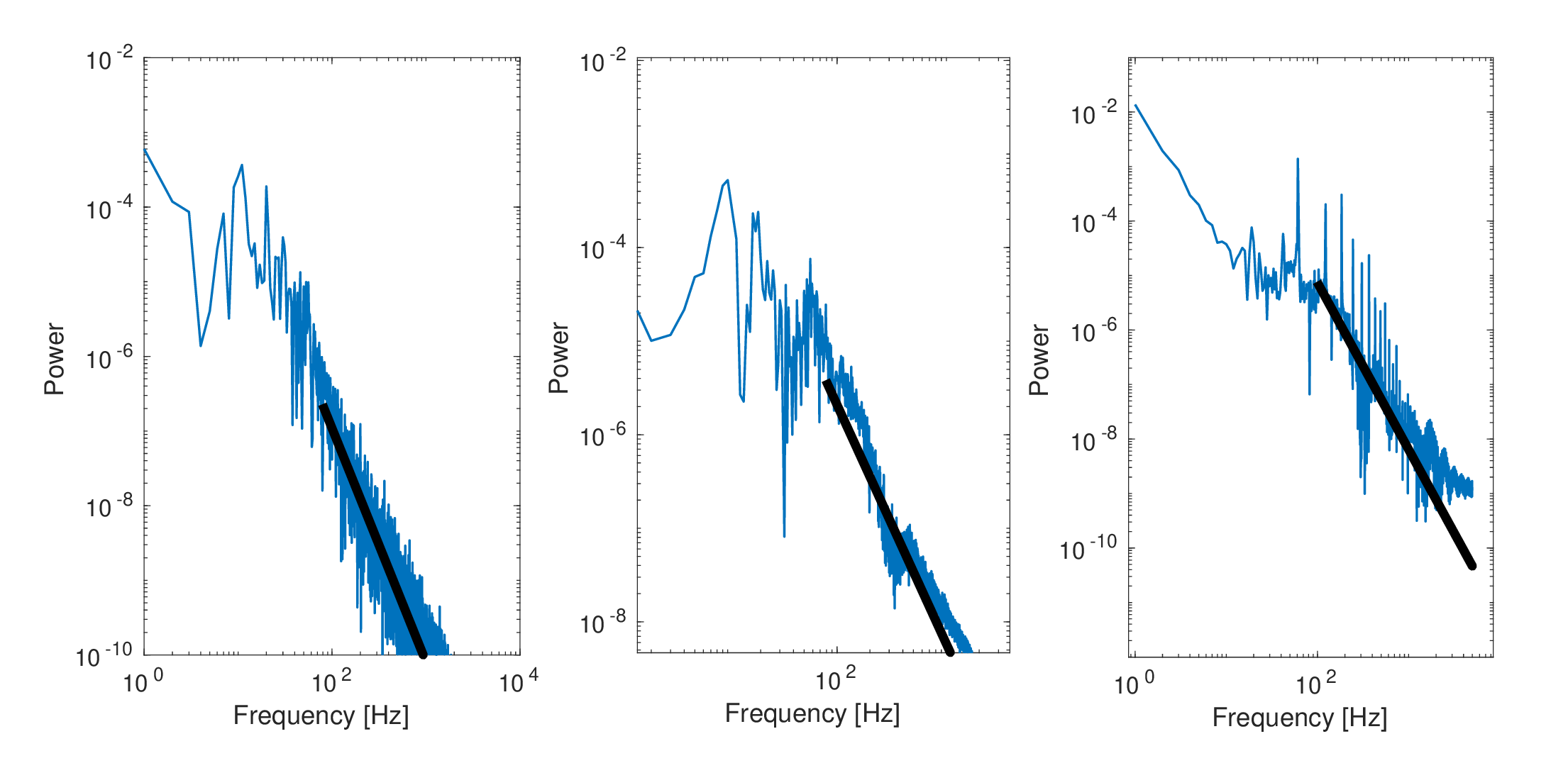}% This is a *.eps file
%\end{picture}
%\includegraphics[width=16cm]{fouriertotsynaptic.eps}% This is a *.eps file
\end{picture}
\end{center}
\caption{Fourier Spectrum of the mean GPi synaptic activity for normal, Parkinsonian and DBS conditions. (a) Normal state, the power shows  2 peaks  at $\approx$7 and 11 Hz representing the mean synaptic variable LFP. Using linear approximation we obtain the slope of the line around -3.6. (b) Fourier Spectrum of the mean GPi synaptic activity under parkinsonian conditions. The power shows a strong lobe at 9Hz and a secondary at 18Hz  reflecting the synchronisation index fluctuations of shown in Fig.\ \ref{fig:Synchidex_allcases}(b) and Fig.\ \ref{fig:MeanSynapGPi}(b). The slope of linear approximation was calculated $\approx$-2.5. (c) Fourier Spectrum of the mean synaptic activity projecting from GPi to THA under STN-DBS conditions. The power peaks at $\approx$60 Hz, with additional sharp peaks between $\approx$ 122 and 184 Hz, i.e., corresponding harmonics, reflecting the mean synaptic activity Fig.\ \ref{fig:MeanSynapGPi}(c) and the  synchronisation index fluctuations shown in Fig.\ \ref{fig:Synchidex_allcases}(c). The slope is changed -3.1 close to normal conditions.}
\label{fig:Fouriertot}            
\end{figure}

%\begin{figure}[h!]
%\begin{center}
%\includegraphics[width=14cm]{gpimeanVDBS.eps}% This is a *.eps file
%\end{center}
%\caption{.}
%\label{fig:Park}            
%\end{figure}

%\begin{figure}[h!]
%\begin{center}
%\includegraphics[width=14cm]{synchGPIdbs.eps}% This is a *.eps file
%\end{center}
%\caption{Synch index during DBS High Frequency.}
%\label{fig:Park}            
%\end{figure}

%\begin{figure}[h!]
%\begin{center}
%\includegraphics[width=13cm]{optfreq.eps}% This is a *.eps file
%\end{center}
%\caption{Optimal frequency for the the to macroscopic indexes. Existence of %three wells with respect the frequency. Regarding the Desynchronization the optimum frequency exists at 150 Hz.}
%\label{fig:Park}            
%\end{figure}

%\begin{figure}[h!]
%\begin{center}
%\includegraphics[width=13cm]{finaldbs.PNG}% This is a *.eps file
%\end{center}
%\caption{Overview on the effects of DBS frequency on synchronisation index, synaptic index, and thalamic %firing in response to cortical input. } 
%\label{fig:ParkDBSopt}            
%\end{figure}

\section{DBS efficiency depends on stimulation frequency}
\label{sec:DBS_Effic}
As the previous considerations show, the macroscopic indices $r$ and $l$, which are defined in eq.\ \eqref{eq:Sync_index} and \eqref{eq:LFP} respectively, allow to describe the effects of DBS stimulation and to compare the different states (normal, parkinsonian, and DBS). 

 One critical parameter of the DBS result is the thalamic response with respect the sensorimotor cortical signal. Faithful thalamic activation will send strong excitatory signals to cortex, see Fig.\ \ref{fig:1direct_inderect}(a), alleviating in this way the parkinsonian symptoms. Naturally, under normal conditions the thalamic response should neither be completely uncoupled (with respect to the sensorimotor cortical signal), as under parkinsonian conditions (see Fig.\ \ref{fig:Park}(d)), nor completely coupled as in an isolated cortico-thalamic system (Fig.\ \ref{fig:4Thalam}). Indeed, under normal conditions, as shown in Fig.\ \ref{fig:Normal}(d), the thalamus follows the cortical input relatively closely, but not in absolute synchrony. 

In order to quantify the thalamic response to cortical input, under normal conditions or DBS with various stimulation frequencies, we define the response efficacy $R$ of thalamic neurons as a mean value of the fraction of activated thalamic neurons during simulated cortical activation. This simulation, as already discussed in Fig.\ \ref{fig:4Thalam}, comprised  sensorimotor current injections of length $\delta$ (5 ms) defined by the interval $[t,t+\delta]$, with frequency of 40 Hz. The mathematical formulation of response efficacy $R$ is thus defined as:
\begin{equation}
    R(f)=\frac{\sum\limits_{i=1}^{N_{\text{int}}} a_i(t) }{N_{\text{int}}} ,
\end{equation}
where $a_i(t)$ is the proportion of activated neurons, i.e. the number of activated neurons within the time interval of $[t,t+2\delta]$ divided by the whole number of thalamic neurons $N_{\text{THA}}=200$. The summation is taken over the number of intervals $N_{\text{int}}$, for times $t>500$ ms, (i.e. $N_{\text{int}}=50$). Under normal conditions, as expected, $R$ is approximately 0.5, i.e. suggesting a coupling to cortical input at a value offering the broadest dynamic range. 

\begin{figure}[h!]
\begin{center}
\includegraphics[width=1\textwidth]{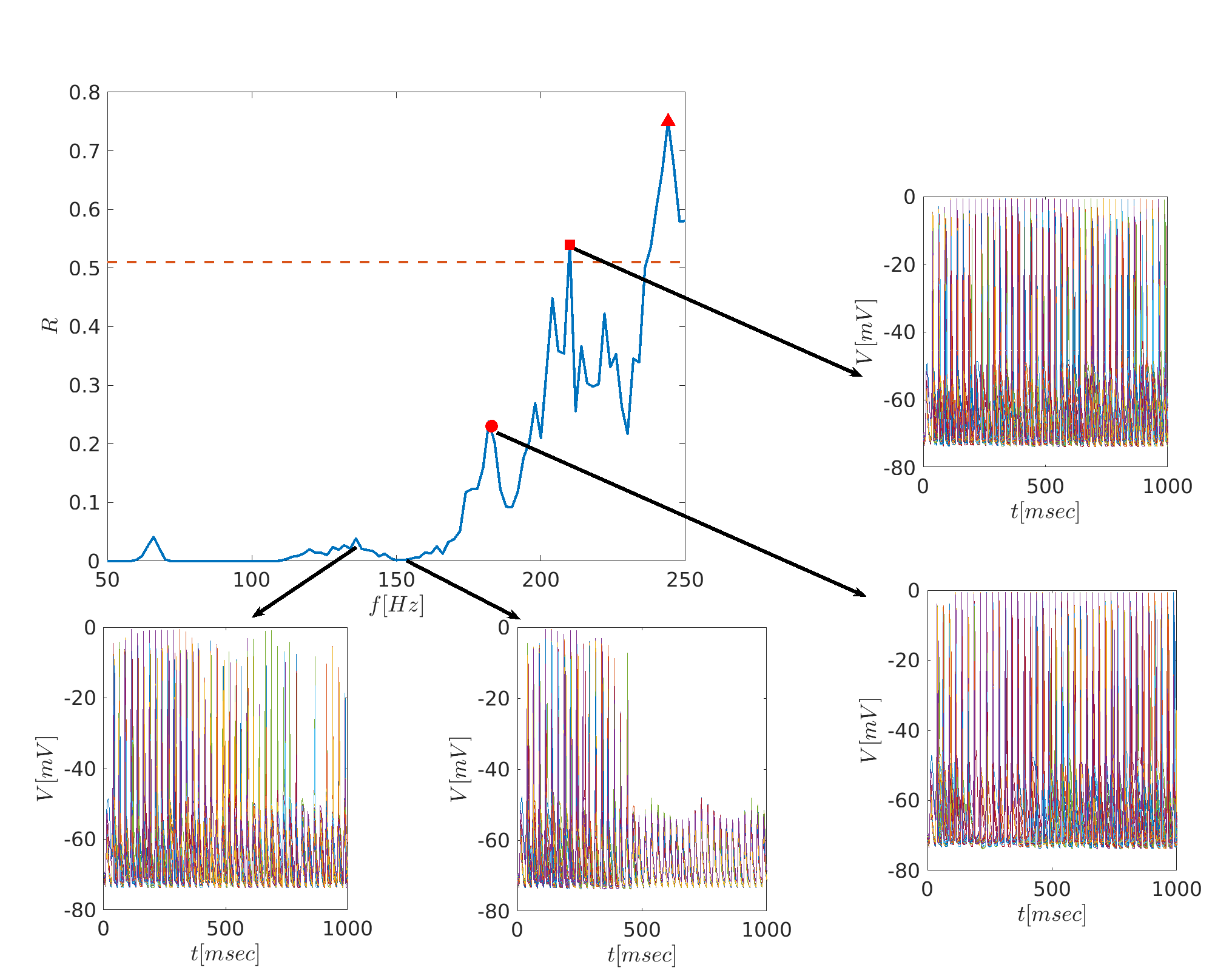}%{maainoptfreq2.eps}% This is a *.eps file
\end{center}
\caption{Response efficacy $R$ of thalamic neurons changes with DBS frequency $f$. The central graph depicts the response efficacy of thalamic neurons to cortical input under DBS. In this context, a value of 1 means thalamic neuronal firing absolutely correlates with cortical input, a value of 0 that there is no correlation whatsoever. Under normal conditions (dashed line), thalamic neuronal firing shows a medium correlation to cortical input, with a response efficacy of 0.518 (compare also Fig.\ \ref{fig:Normal}(d), whereas under parkinsonian conditions, this correlation is almost completely lost, with response efficacy of close to 0 (compare Fig.\ \ref{fig:Park}(d). In our modelling study, using 184 Hz as DBS frequency, the response efficacy is close normal (i.e. 0.28; see first peak of the curve highlighted by filled circle). The inseones show the resulting firing behaviour of thalamic neurons in response to cortical input at the different frequencies $f$ as indicated. Note that at 150 Hz, the impact of DBS is only transient, lasting only up to 500 ms.} 
%Three phase behaviour. In LFS no effective stimulus. Second phase after 150 until 184 HZ and then from 216 up til 234 HZ. Effective DBS should be at 184 and 234 Hz}
\label{fig:Effic}            
\end{figure}

Under DBS at various frequencies, $R$ shows a non-linear behaviour, starting at 0 for low DBS frequencies (around 50 Hz) to values $>0$ with small peaks for frequencies around 70 and 130 Hz. Very prominent peaks are found around 184 Hz, 210 Hz and 244 Hz. Thus, there are three dominant optima at  where the DBS maximises the thalamic response close to normal values of $R$. As Fig.\ \ref{fig:DPS_T} thus shows, the frequency of DBS is critical for thalamic firing outcome. At low frequencies (i.e. 50-150 Hz), only few frequency bands can be found at which $R > 0$. Consequently, the effects on thalamic activation are only transient at e.g. 50 Hz and 150 Hz (see Fig.\ \ref{fig:Effic}). Exceptions are the small peaks at 70 and 130 Hz; Fig.\ \ref{fig:Effic}, reasonably stable responses can thus be seen at 130 Hz.  Only from 160 Hz onward, stable thalamic firing can be achieved by DBS. Interestingly, this is very similar to experimental findings in hemiparkinsonian rats: low frequency DBS stimulation up to 75 Hz actually result only in transient effects, while permanent reductions of circling behaviour (the parkinsonian equivalent in this animal model) were only achieved at DBS frequencies $>$ 130 Hz \cite{so17}.  

Computing the Shannon entropy for the macroscopic variables $r,l$ we confirm the optimal DBS frequency.  As suggested by \cite{Dorv08,Arle18,Deco12}, an optimised  DBS frequency can be achieved by regularisation of the whole basal ganglia activity\cite{Dorv08}, i.e. minimisation of the entropy and thus a more ordered state.

\begin{figure}[h!]
\begin{center}
\includegraphics[width=.8\textwidth]{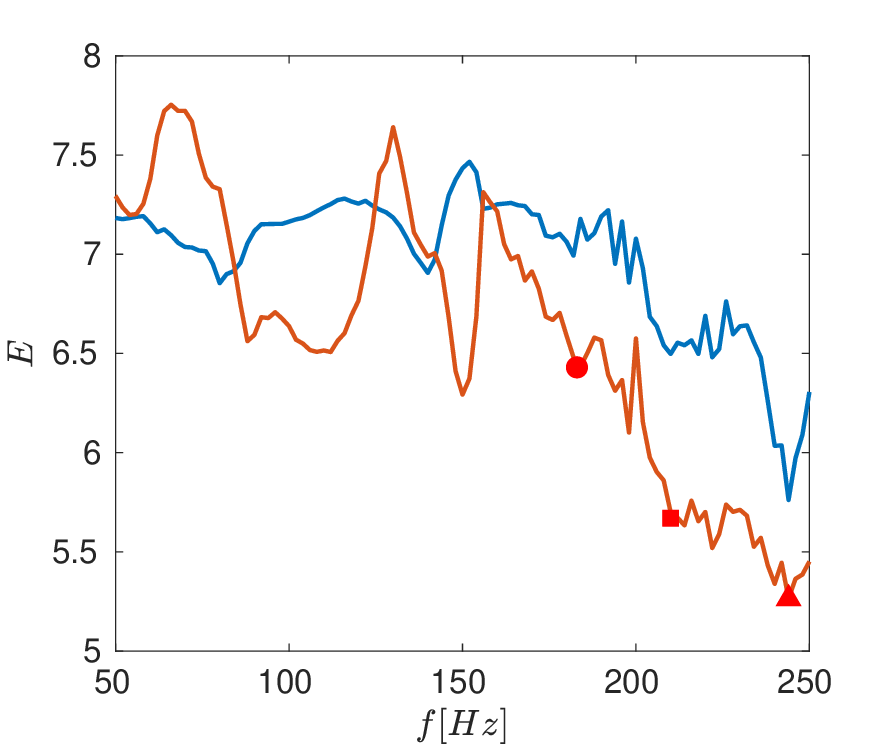}%{entropy.eps}% This is a *.eps file
\end{center}
\caption{Entropy of synchronisation and synaptic indices deceases with increasing DBS frequency. The graph shows both single entropy values referring to (i) the  synchronisation index and to (ii) mean synaptic GPi activity index marked with (with markers). The figure is in correspondence with response efficacy, see Fig.\ \ref{fig:Effic} with local minima at 184, 210 and 244 Hz marked with circle, square and triangle respectively.}
\label{fig:Entropy}            
\end{figure}

We thus compute the Shannon Entropy $E$, by 
\begin{equation}
    E_x(f)=-\sum_{i=0}^{N}P_f(x_i)\ln P_f(x_i),
\end{equation}
where $x_i$ express the macroscopic variable of interest (here $x=r, l$). Calculating the entropy for different frequencies, we obtain 
Fig.\ \ref{fig:Entropy}. As can be deduced from this figure, one DBS optimum
frequencies  indeed be at a frequency $f$ of 184, 210 Hz, 
where entropy values of both synchronisation and synaptic GPi 
activities reach a local minimum (and also, thalamic 
response efficacy $R$ is highest, cf.\ Fig.\ \ref{fig:Effic}).
A similar minimum can be found when increasing $f$ further 
at 244 Hz, which would require more 
energy for the DBS stimulation, likely raising the possibility of 
energy-dependent side effects. 
%\textcolor{red}{Why is this the case? We should give arguments to explain this.}

\section{Discussion and Conclusions}
%The Entropy was calculated for the the two macroscopic variables the synchronisation index and the mean synaptic activity. Fig.\ \ref{fig:Entropy}
     
A large scale basal ganglia computational model has been used to study network dynamics in movement disorders. The network model consists of 1700 interactive neurons with approximately 30000 connections (representing the \emph{microscopic} level). For different values of parameters, the model reproduces phase transitions for normal, parkinsonian and DBS, in the emergent dynamics which are captured with suitable \emph{macroscopic} indices (for example the change from oscillatory to bursting behaviour of mean synaptic activity in Fig.\ \ref{fig:MeanSynapGPi}(a)and(b)). Importantly, transitions produced by the model are consistent  with the physiology and experimental observations of aberrant functionality of direct and indirect pathways.\\
Experimental findings of animal studies, regarding the impact of STN-DBS on GPe, GPi and thalamic firing could closely be replicated (qualitatively). Specifically, in the transition from normal to parkinsonian state shown in  Figs.\ \ref{fig:Normal} and \ref{fig:Park}, the model alters the dynamics of GPe and GPi neurons due to varying levels of striatal inhibition. Remarkably, similar alternations in GPe/GPi dynamics were observed in \cite{Galv08} (Fig.\ 2 therein) in monkeys and mice treated with MPTP (methyl-4-phenyl-1,2,3,6-tetrahydropyridine)\cite{Galv08,Oga85}. 
%Both studies conclude that the parkinsonian state emerges as a complex network disorder, in which abnormal activity in groups of neurons in the basal ganglia  which strongly affects the excitability, oscillatory activity, synchrony and sensory responses of areas of the cerebral cortex related to planning and execution of movement \cite{Galv08}. 
Furthermore, under DBS, in the model, the GPi and GPe fire tonically at high frequency, following STN firing at stimulation frequency (184 Hz), abolishing synchronised $\beta$ activity. This corresponds to the findings of \cite{McCon12}, who showed that STN stimulation results in a sharp power peak at the same frequency also in GPe, and with findings of Wang et al. \cite{Wang18} where DBS disrupts pallidal beta oscillations and the cortical coherence in Parkinson disease. 

Regarding the $\beta$-band hypersynchrony hypothesis of Parkinson's disease \cite{Kuhn08}, the model thus importantly shows that DBS actually abolishes $\beta$-band synchrony, which also in the model is prevalent in GP without stimulation. Hence, the model faithfully replicates what is known from clinical and animal model studies both on the cellular and on the network activity pattern levels. The model produces clusters of local travelling waves, more pronounced in the parkinsonian state. Remarkably, in the experimental findings of \cite{Cag15}, it has been found that the analysis of local field potential recordings from the subthalamic nucleus and globus pallidus of patients with Parkinson's disease show beta band propagation waves within the globus pallidus.

Further, the model predicts that a faster firing GPi projections under DBS will result in regular firing of the thalamus, which in turn will essentially follow the cortical input - while under PD conditions, in fact thalamic firing was very sparse and irregular, and did not faithfully mirror cortical signals. These results are supported by studies in a parkinsonian animal model in rhesus monkeys \cite{Xu08}, where STN-DBS (as in Fig.\ \ref{fig:DPS_T}) produces a change in the pattern and periodicity of neuronal activity in the basal ganglia thalamic network, resulting in a regular, higher-frequency firing pattern in the thalamus.

Moreover, macroscopic properties can be derived from our model. Since these essentially mirror local field potential activity, these properties can be used to test predictive modelling in future studies. For example in \cite{HE10} similar changes in the critical exponent \emph{a} (which characterises the distribution of power spectrum) were found. They explain these differences by different areas of activation during the experiments. In our case, this exponent is the slope of the linear approximation in Fig.\ \ref{fig:Fouriertot}. Our analysis shows that the slope can be distinguish between normal parkinsonian and DBS state. Further, the model also allows to extract variables such as a macroscopic synchronisation and GPi synaptic activity indices which theoretically allow for predicting the probability of transition into parkinsonian state, which might pave the way for feedback control of DBS in the future \cite{Popo18,Popo19,Manos18}.

Beyond this, the detailed analysis of the macroscopic parameters, and derived values such as entropy also allow for an optimisation of DBS. As an outcome measure, the response efficacy of thalamic neurons reflects the degree of thalamic activity correlating to cortical input, which under normal conditions is $\approx$ 0.5. Critically, raising stimulation frequency beginning with 50 Hz, the first prominent entropy local  minima of the two mesoscopic parameters precisely coincide with the first peak of the response efficacy $R$ close to normal values ($\approx$ 0.23) at 184 Hz and more efficient frequencies over the 200Hz. Although these frequencies are not identical to the  most commonly used one in clinical settings (i.e.\ 130 Hz), the optimum obtained in this study is close. More importantly, the model, and more specifically the response efficacy $R$ qualitatively very much reproduces data from animal studies in hemiparkinsonian rats \cite{so17}, where clinical improvement was seen only with increasing DBS frequencies from 75 Hz onward, with two optima at 130 and 260 Hz, and less efficacious frequencies in between. Our results confirm the experimental findings of \cite{so17} showing that DBS above 130Hz is more effective in hemiparkinsonian rats. Our mathematical model extends previous computational work \cite{So12,Dorv10} by using a larger number of neurons for each basal ganglia areas and complex small world  connectivity. Additionally, we propose different approximations in the DBS frequency. The frequency analysis of \cite{Dorv10} differs from our investigation since they focus on perturbations around 130 Hz (Fig. 1 and 2 in \cite{Dorv10}). In \cite{So12} the activation patterns of local cells and fibers passage are studied with respect the fidelity of thalamus. Our model suggests that the basal ganglia network behaviour to DBS stimulate frequency has a strong nonlinear response similar to resonating mechanical systems with optimal frequencies above 130Hz, suggesting the investigation of DBS treatment beyond the 130Hz. In summary, the mathematical model and the analysis is considered a powerful tool to explore the effects of parameter-dependent changes of DBS and to optimise the medical treatment.

%\begin{figure}[h!]
%\begin{center}
%\includegraphics[width=14cm]{travelling_breather%GS.jpg}% This is a *.eps file
%\end{center}
%\caption{Similar to Gray-Scott model for %breathers. Taken from ''Far from Equilibrium Dyanmics'' Nishiura page 259 shows saddle-node bifurcation mechanism controls the onset of splitting.}
%\label{fig:TravelGRay}            
%\end{figure}

\section{Outlook}

%Building on this effort, the computational Basal Ganglia model allows us to investigate several open problems concerning the development of movement disorders. 
One major future topic will be the study of network topological variations and the impact on the emergent dynamics. Thus, the functional effects of neuro-anatomical changes observed in Parkinson's disease \cite{Prak16}, such as massive decreases of dendritic length of medium spiny neurons \cite{Step05}, could be modelled. 
%which showed a reduction of numerical density of dendritic spines on dendrites (about 27$\%$ in striatum, concluding that is likely to have a grave impact on the ability of these neurones to function normally resulting to sympto-pathology of PD. 
Considering movement disorders beyond Parkinsons's disease, the model could also be extended to further elements of the basal ganglia, to gauge the effect of alterations in cortico-striatal communication such as those observed in dystonic hamsters \cite{Kohl04}, where alternations in excitability in dystonic tissue was described to be related to both changes in intrinsic neuronal properties, and presynaptic release probability at glutamatergic synapses.\\
In future work, it will be important to investigate the parameter dependence of network dynamics, using numerical bifurcation tools for multiscale problems \cite{Spil11,Mars14,Moon15,Schm18}, for an in-depth understanding of the functional network changes occurring in movement disorders.

\section*{Funding}
The authors thank the DFG for support through the Collaborative Research Center CRC 1270 (Deutsche Forschungsgemeinschaft, Grant/ Award Number: SFB 1270/1–299150580).

\section*{Acknowledgments}
KS would like to acknowledge George Georgiou and Constantinos Siettos for their help, discussions and encouragement during the early stages of this research. 
%RK would like to acknowledge Angelika Richter for the long-standing collaboration in experimental dystonia research. 
All authors thank Susann Dittmer for her help drawing Fig.\ \ref{fig:1direct_inderect}. 
\begin{table}[ht]
\caption{The following table depicts the values of parameters that used in section 2 for mathematical modelling.}
\label{tab:1}       % Give a unique label
 %For LaTeX tables use
\begin{tabular}{llll}
\hline\noalign{\smallskip}
STN & value  & GPe/GPi & value \\
\noalign{\smallskip}\hline\noalign{\smallskip}
$g_{\text{LEAK}}$ &2.25 nS/$\mu m^2$ & $g_{\text{LEAK}}$  & 0.1 nS/$\mu m^2$ \\
$g_{\text{K}}$ & 45.0 nS/$\mu m^2$ & $g_{\text{K}}$  & 30 nS/$\mu m^2$ \\
$g_{\text{Na}}$ & 37.5 nS $\mu m^2$ & $g_{\text{Na}}$  & 120 nS/$\mu m^2$\\
$g_{\text{T}}$ & 0.5 nS/$\mu m^2$ &$g_{\text{T}}$ & 0.5 nS/$\mu m^2$ \\
$g_{\text{Ca}}$  & 0.5 nS/$\mu m^2$&  $g_{\text{Ca}}$   & 0.15 nS/$\mu m^2$ \\
$g_{\text{AHP}}$ & 9.0 nS/$\mu m^2$ &  $g_{\text{AHP}}$  & 30.0 nS/$\mu m^2$ \\
$E_{\text{LEAK}}$  & -60.0 mV & $E_{\text{LEAK}}$ & -55.0 mV \\
$E_{\text{K}}$& -80.0 mV& $E_{\text{K}}$ & -80.0 mV\\
$E_{\text{Na}}$ & 55.0 mV & $E_{\text{Na}}$ & 55.0 mV \\
$E_{\text{Ca}}$ & 140.0 mV & $E_{\text{L}}$ & 120.0 mV \\
$\tau_{h1}$ & 500.0 msec &  $\tau_{h1}$ & 0.27 msec \\
$\tau_{n1}$ & 100.0 msec & $\tau_{n1}$ & 0.27 msec \\
$\tau_{r1}$  & 17.5 msec & $\tau_{r1}$ & 0.05 msec \\
$\tau_{h0}$ & 1.0 msec & $\tau_{h0}$  & .05 msec \\
$\tau_{n0}$ & 1.0 msec & $\tau_{n0}$ & 1.0 msec \\
$\tau_{r0}$ & 40.0 msec & $\tau_{r0}$  & 30.0 msec \\
$k_1$ &15.0 &$k_1$ &30.0 \\
$k_{\text{Ca}}$ & 22.5 & $k_{\text{Ca}}$ & 20 \\
$k_2$ & $3.75 \cdot 10^{-5}$  $\text{msec}^{-1}$ & $k_2$ & $ 10^{-4}\text{msec}^{-1}$ \\
$\theta_m$ & -30.0 & $\theta_m$ & -37 \\
$\theta _h$ & -39.0 & $\theta _h $ & -58 \\
$\theta _n$ & -32.0 & $\theta _n $ & -50 \\
$\theta _r$  & -67.0 & $\theta _r $ & -70\\	 
$\theta _a$ & -63.0 & $\theta _a $ & 63.0 \\
$\theta _b $ & 0.4 & $\theta _b $ & 0.4 \\
$\theta _s $ & -39.0 & $\theta _s $ & 39.0\\
$\theta _{\tau h }$ & -57.0 & $\theta _{\tau h }$ & -40.0\\
$\theta _{\tau n }$ & -80.0 & $\theta _{\tau n }$ & -40.0\\
$\theta _{\tau r }$ & 68.0 & $\theta _{\tau r }$ & -\\
$\sigma_m$ & 15 & $\sigma_m$ & 10 \\
$\sigma_h$ & -31 & $\sigma_h$ & -12 \\
$\sigma_n$ & 8 & $\sigma_n$ & 14 \\
$\sigma_r$ & -2 & $\sigma_r$ & -2 \\
$\sigma_a$ & 7.8 & $\sigma_a$ & 2 \\
$\sigma_b$ & -0.1 & $\sigma_b$ & - \\
$\sigma_{\tau h }$ & -3 & $\sigma_{\tau h }$ & -37 \\
$\sigma_{\tau n }$ & -26 & $\sigma_{\tau n }$ & -37 \\
$\sigma_{\tau r }$ & -2.2 & $\sigma_{\tau r }$ & - \\
$A_h$ & 0.75 & $A_h$ & 0.05 \\
$A_n$ & 0.75 & $A_n$ & 0.05 \\
$A_r$ & 0.12 & $A_r$ & 2 \\
$\alpha $ & 5 & $\alpha$ & 2 \\
$\beta $ & 1 & $\beta $ & 0.08 \\
$\theta_0 $ & -39 & $\theta_0 $ & -57 \\
$A_{\text{DBS}}$ & 200 & - &  - \\
$\delta_{\text{DBS}}$ & .6 msec & - &  - \\
$T_{\text{DBS}}$ & 6 msec & - &  - \\
\noalign{\smallskip}\hline
\end{tabular}
\end{table}

\begin{table}
\caption{The values of parameters that used for THA are given in the next table.}
\label{tab:2}       % Give a unique label
 %For LaTeX tables use
\begin{tabular}{ll}
\hline\noalign{\smallskip}
THA & value  \\
\noalign{\smallskip}\hline\noalign{\smallskip}
$g_{\text{LEAK}}$ &0.05 nS/$\mu m^2$ \\
$g_{\text{K}}$ & 5 nS/$\mu m^2$ \\
$g_{\text{Na}}$ & 3 nS $\mu m^2$ \\
$g_{\text{T}}$ & 5 nS/$\mu m^2$ \\
$g_{\text{Ca}}$  & 0.5 nS/$\mu m^2$\\
$E_{\text{L}}$  & -70.0 mV \\
$E_{\text{K}}$& -90.0 mV \\
$E_{\text{Na}}$ & 50.0 mV \\
$E_{\text{Ca}}$ & 140.0 mV \\
$E_{\text{T}}$  & 0 mV \\
$\tau_{h1}$ & 500.0 msec  \\
$\tau_{n1}$ & 100.0 msec \\
$\tau_{r1}$  & 17.5 msec \\
$\tau_{h0}$ & 1.0 msec \\
$\tau_{r0}$ & 40.0 msec \\
$k_1$ &15.0 \\
$\theta _h$ & -41.0 \\
$\theta _r$ & -84.0 \\
$\theta _m$ & -37.0 \\
$\theta _p$ & -60.0 \\
$\sigma_h$ & 4 \\
$\sigma_r$ & 4\\
$\sigma_m$ & 7\\
$\sigma_m$ & 6.2\\
$A_{\text{SM}}$ & 5 \\
$\delta_{\text{SM}}$ & 5 msec\\
$T_{\text{SM}}$ & 25 msec \\
\noalign{\smallskip}\hline
\end{tabular}
\end{table}

%\begin{acknowledgements}
%\end{acknowledgements}


\bibliographystyle{spmpsci}      % mathematics and physical sciences
%\begin{thebibliography}{}
% etc
%\end{thebibliography}
%\printbibliography
\bibliography{bibl}
\end{document}